\documentclass[preprint,3p, authoryear]{elsarticle}

\usepackage{dblfnote}
\usepackage{graphicx}
\usepackage{blindtext}
\usepackage[inline]{enumitem}
\usepackage{amssymb}
\usepackage{xspace}
\usepackage{amsmath}
\usepackage{lineno,hyperref}
\usepackage{algorithm}
\usepackage{algorithmic}
\usepackage{tabu}
\usepackage{rotating}
\usepackage[lofdepth,lotdepth]{subfig}
\usepackage{tikz}
\def\checkmark{\tikz\fill[scale=0.4](0,.35) -- (.25,0) -- (1,.7) -- (.25,.15) -- cycle;}

\journal{Journal of Expert Systems With Applications}

\makeatletter
\long\def\MaketitleBox{%
	\resetTitleCounters
	\def\baselinestretch{1}%
	\begin{center}%
		\def\baselinestretch{1}%
		\Large\@title\par\vskip18pt
		\normalsize\elsauthors\par\vskip10pt
		\footnotesize\itshape\elsaddress\par\vskip36pt
	\end{center}%
}
\makeatother

\begin{document}
	
\begin{frontmatter}
\title{ExEm: Expert Embedding using dominating set theory with deep learning approaches}

\author{Narjes Nikzad-Khasmakhi}
\address{Department of Computer Engineering, University of Tabriz, Tabriz, Iran\\
n.nikzad@tabrizu.ac.ir}
\author{Mohammadali Balafar \corref{mycorrespondingauthor}}
\cortext[mycorrespondingauthor]{Corresponding author}
\address{Department of Computer Engineering, University of Tabriz, Tabriz, Iran\\
balafarila@tabrizu.ac.ir}
\author{M.\, Reza Feizi-Derakhshi}
\address{Department of Computer Engineering, University of Tabriz, Tabriz, Iran\\
mfeizi@tabrizu.ac.ir}
\author{Cina Motamed}
\address{Department of Computer Science, University of Orléans, Orléans, France\\
motamed@free.fr}

\end{frontmatter}

\newpage
\textbf{Abstract}
	A collaborative network is a social network that is comprised of experts who cooperate with each other to fulfill a special goal. Analyzing this network yields meaningful information about the expertise of these experts and their subject areas. To perform the analysis, graph embedding techniques have emerged as an effective and promising tool. Graph embedding attempts to represent graph nodes as low-dimensional vectors. In this paper, we propose a graph embedding method, called \textit{ExEm}, that uses dominating-set theory and deep learning approaches to capture node representations. ExEm finds dominating nodes of the collaborative  network and constructs intelligent random walks that comprise of at least two dominating nodes. One dominating node should appear at the beginning of each path sampled to characterize the local neighborhoods. Moreover, the second dominating node reflects the global structure information. To learn the node embeddings, ExEm exploits three embedding methods including Word2vec, fastText and the concatenation of these two. The final result is the low-dimensional vectors of experts, called expert embeddings. The extracted expert embeddings can be applied to many applications. In order to extend these embeddings into the expert recommendation system, we introduce a novel strategy that uses expert vectors to calculate experts' scores and recommend experts. At the end, we conduct extensive experiments to validate the effectiveness of ExEm through assessing its performance over multi-label classification, link prediction, and recommendation tasks on common datasets and our collected data formed by crawling the vast author Scopus profiles. The experiments show that ExEm outperforms the baselines especially in dense networks. 

\textbf{keyword}
	Social and collaborative Networks,  Graph embedding, Node representation, Dominating set, Expert recommendation system   
	


\section{Introduction}\label{sec:introduction}
Social networks have emerged as a great platform to generate and share information. Social networks consist of entities and the interactions between them. A common representation for the social network has nodes for entities, and edges linking two nodes to denote the relationships between entities \citep{POWELL2015111}. 

A collaborative network is observed as a specific type of social networks that is comprised of experts who cooperate with each other to fulfill a special goal. Analyzing this network yields meaningful information about the expertise of these experts and their subject areas. Although a collaborative network provides a rich source of information about experts, a major challenge surrounding this network is how to analyze both its structure and content. By the way of illustration, Question Answering Community (\textit{QAC}) is one types of collaborative network in which the users' collaborations are asking or answering questions \citep{Zhao:2016:EFC:3060832.3061041}.  One of the key problems in QAC is how to find users, called experts, to answer the given questions \citep{NIKZADKHASMAKHI2019126}. An ordinary solution to this issue is analyzing the social interactions of users and content of the questions asked and the answers replied by them \citep{WANG20131442}. As another example, academic papers are composed of several co-authors. The development of cooperation among academic authors constitutes a collaborative network, called the co-author network where the connections  demonstrate the corresponding authors have published at least one paper together \citep{li2017interpreting}. In the co-author network, analyzing the interactions of authors and the content of their papers have created a way to recognize the key researchers in a specific area, who are defined as experts \citep{liu2015new}. 

Representing data in particular social networks in the form of graphs has been attracting increasing attention in recent years. On the other hand, performing analysis on this type of data structure helps us gain more information and insights. Graph analytic appears in a wide variety of applications such as node classification, link prediction, clustering, node recommendation, visualization, and etc \citep{cai2018comprehensive, DBLP:journals/corr/GoyalF17}. Although many methods have been proposed for the graph representation and analysis, they encounter several challenges such as memory cost and time complexity. Graph embedding is an effective solution that tries to overcome these issues. It changes the form of representing a graph and maps the nodes into a low-dimensional space. Also,  it's able to maintain consistent structural information and properties of the graph.  

As mentioned before,  applying a graph embedding method on the resulting graph of a social network creates a better understanding of the  network  entities and its structure \citep{KEIKHA201847}. By taking a glimpse of the previous graph embedding techniques, it's obvious that a group of proposed approaches denote a graph as a set of random walks which preserve the graph characteristics \citep{cai2018comprehensive}. After that,  the deep learning methods  such as skip-gram are applied to the random walks to maximize the likelihood of observing neighborhoods of nodes.  The crucial difference between these methods is the way of generating random walks. DeepWalk \citep{perozzi2014deepwalk} and Node2vec \citep{grover2016Node2vec} are two examples of this category. Although DeepWalk uses a simple depth-first search process for making random walks, it suffers from the repeated nodes problem and does not consider the efficacy of breadth-first neighbor structure. On the other hand, Node2vec develops biased-random walks using the breadth-first and depth-first search strategies. Node2vec has two parameters $P$ and $Q$ that help control over the search space. One of the drawbacks of Node2vec is the necessity to always set the outperform values for these parameters for every network \citep{gu2018graphlets}.

In this research, we propose a deep learning graph embedding with random walk that is called \textit{ExEm}. We aim to transform a graph into a low-dimensional vector space using dominating nodes in creating random walks. We also investigate the effect of modified random walks on the quality of produced  node embeddings. Dominating set theory has been a classic subject studied in graph theory that is considered as a virtual backbone in these areas \citep{du2012connected}.  A set is dominating if every node in the network is either in the set or a neighbor of a node in this set \citep{1652947}. ExEm generates a set of random walks that satisfy two conditions: starting random walk with a dominating node and containing at least another dominating node in the sampled path.  The existence of dominating nodes in the sampled path enables ExEm to capture the local and global network structures. In short, the dominating set is an approximation of the network that manages the entire network \citep{sun2019minimum}. Hence, these intelligent random walks are the main cause in learning rich feature representation for the graph. After producing the desired random walks, they are stored as the sentences of a corpus. Then, skip-gram neural network model is used to map the nodes into the embedding vectors. We train this neural network by Word2vec and fastText. Also, we consider another node representation which is the combination of the embeddings extracted from Word2vec and fastText. Moreover, the effectiveness of graph embedding in different real-world applications motivates us to explore its potential usage in the expert recommendation system and proposes a strategy to compute experts' scores and recommends experts. On the other hand,  we present a collaborative network that is constructed based on the gathered information from \textit{Scopus}.  In this network, nodes with multi labels are represented as authors. The  node labels demonstrate the author' subject areas. Edges between authors denote their co-author relationship.

\textbf{Research Questions}: we aim to answer the following research questions in this study:

\begin{enumerate}[label=\textbf{RQ.\arabic*},ref=RQ.\arabic*, leftmargin=2cm]
	\item \label{rq1} Does the data gathered from Scopus provide a suitable real dataset for the different tasks such as classification, link prediction, recommendation and so on?
	\item \label{rq2} How does using dominating set theory affect the performance of node representation learning?
	\item \label{rq3} How can we extend the obtained node representations into expert recommendation systems to recommend experts?
\end{enumerate}

The remainder of the paper is outlined as follows: Section \ref{related_work} reviews the related works. Section \ref{Proposed_Method} explains our proposed method in detail. Section \ref{data_dis} presents the descriptions of the gathered dataset from Scopus. In order to verify the proposed approach, extensive experiments are conducted on real-life datasets. The descriptions of these datasets and baseline approaches, parameter setting and the evaluation metrics used to test our proposed method are presented in Section \ref{sec:experimentalEvaluation}.  The experimental results and their analysis are given in Section \ref{evaluation_re}. Section \ref{discussion} discusses the test results.  Section \ref{ans-que} answers the research questions. Finally, Section \ref{sec:conclusion} concludes the paper.

\section{Related Work}\label{related_work}
To analyze social network data, previous studies have proven that the representation of social network as a graph structure and using graph theories have achieved successful results. On the other hand, deep learning based approaches have been demonstrated to be a promising technique to analyze information from social networks with complicated structures. Hence, the incorporation of graph-structured data and the deep learning models results in an outstanding feature learning technique, called graph embedding. Graph embedding learns a map of the graph's nodes to a low-dimensional space features. It provides insight into analyzing users' activity patterns and their relationships in social networks. In this section, we investigate some of the proposed graph embedding methods by different researches.

GraRep \citep{cao2015grarep} learns the node representations of weighted graphs. It uses the matrix factorization version of skip-gram to obtain high-order proximity \citep{cai2018comprehensive}. On the other hand, it catches the $k$-step $(k = 1, 2, 3, ...)$ neighbour relations and integrates global structural information of the graph into the learning process. The final representation of nodes are provided by concatenating $k$-step node representations together \citep{cui2018survey, zhang2018network}.

TriDNR \citep{pan2016tri} utilizes the structure, content, and labels of nodes for constructing the graph embedding. It learns the network structure by the help of DeepWalk approach. Moreover, TriDNR couples two neural networks to capture the node content and label information. Finally, the obtained representations from network structure, and the node label and attribute are linearly combined together \citep{liao2018attributed,zhang2018network,cai2018comprehensive}.

Mahmood et al. \citep{mahmood2016using} have proposed a geodesic density gradient (GDG) algorithm that is divided a network into a series of relatively small communities \citep{wang2019detecting, ahuja2018finding}. This study considers a vector for each node with dimensionality equals the number of all nodes. In this vector, every dimension represents the geodesic distance of that node from all other network nodes  \citep{cai2018comprehensive}. Thus, the network structure can be captured from the geodesic distance vectors. In this way, the nodes with the same region of space belong to the same communities in the original network.

DNGR \citep{cao2016deep} is based on deep learning that aims to construct the low-dimensional vector representations from the PPMI matrix. To achieve this target, DNGR comprises of three steps. At the first step, it obtains information related to the graph structure by proposing a random surfing model which is inspired by the PageRank model. Then, DNGR creates a probabilistic co-occurrence matrix. Subsequently, the PPMI matrix is built based on the probabilistic co-occurrence matrix. Finally, a stacked denoising auto-encoder is applied to the PPMI matrix to learn the embeddings \citep{cui2018survey}.

HOPE \citep{Ou:2016:ATP:2939672.2939751} is a matrix factorization based method. It captures the asymmetric transitivity property of a directed network in embedding vectors \citep{cui2018survey}.  Asymmetric transitivity describes the correlation among directed edges. HOPE measures the high-order proximity from four measurements including Katz Index, Rooted Page Rank, Common Neighbors, and Adamic-Adar score. Then, a generalized Singular Value Decomposition (SVD) is applied to the the high-order proximity to obtain the low-dimensional representations \citep{DBLP:journals/corr/GoyalF17, zhang2018network}.

Although many network embedding methods are proposed for static networks, recent attempts have investigated the embedding methods over the dynamic networks that evolve over time \citep{zhu2018high,mahdavi2018dynNode2vec,taheri2019learning}.  Goyal et al. \citep{goyal2019dyngraph2vec} recommends a deep learning model to capture temporal patterns in dynamic networks for the link prediction task.  This study introduces three different architectures using an auto-encoder, LSTM, and combination of these both.  These architectures take as input the adjacent matrix $A_{t-l}\left[i\right]$, $A_{t-l+1}\left[i\right]$, $\dots$, $A_{t-1}\left[i\right]$ and produce a vector $v_{t_i}$ corresponding to the embedding of  $v_{i}$ at time $t$. They allow predicting interactions between vertices at each time step. Moreover, in another study \citep{sankar2018dynamic}, the authors propose to compute a dynamic node representation by employing self-attention mechanisms over its neighbors and previous historical representations. The survey \citep{kazemi2019relational} reviews the recent representation learning methods for dynamic graphs.

Additionally, some studies have focused on the knowledge graph embedding. A knowledge graph is a directed graph that represents structured information of entities as nodes and their relations as edges \citep{huang2019knowledge, zhang2018network}. The research \citep{guo2016sse} embeds the knowledge graph in this manner that entities are closed to each other in the embedding space if they belong to the same semantic category. Authors in \citep{wang2017knowledge} provides a review of existing approaches presented for knowledge graph embedding.

In spite of the fact that graph embedding is a powerful tool for converting  graph data into low dimensions, employing all features for this purpose may lead to noise \citep{chen2014unified}. To handle this challenge, one solution is dimensionality reduction. In recent years, many studies have focused on the usage of dimensionality reduction for graph embedding. Dimensionality reduction methods are categorized into two groups: feature selection and feature extraction \citep{ZHU2019458}.  Chen et al. \citep{chen2014unified} proposes a binary feature selector by exploiting the least squares formulation of graph embedding. The paper \citep{nishana2013graph} conducts a discussion of the most popular linear dimensionality reduction methods. 

Moreover, a number of surveys have been conducted to categorize the existing graph embedding methods based on their proposed techniques. Cai et al.  \citep{cai2018comprehensive} summarizes the researches into five categories: matrix factorization, deep learning, edge reconstruction, graph kernel, and generative model. In this study, deep learning based graph embedding is divided into two groups, deep learning graph embedding with and without a random walk.  Based on the viewpoint of this review, an edge reconstructing based graph embedding technique minimizes the distance loss to preserve first- and second-order proximities. On the other hand, a matrix factorization based method represents the connections between nodes as a matrix and factorizes this matrix to extract node embedding.  Moreover, deep learning based graph embedding techniques with random walks represent a graph as a set of random walks and these random walks are fed into a deep learning method like skip-gram to optimize their neighborhood preserving likelihood objectives. In comparison, deep learning based graph embedding methods without random walks apply deep neural networks such as auto-encoders or convolutional neural network, on the whole graph. Zhang et al.  \citep{zhang2018network} reviewes the state-of-art graph embedding techniques with a different outlook. They classify the studies into two classes: unsupervised and semi-supervised network representation learning. Also, this survey summarizes the existing approaches from methodology perspective into five types: matrix factorization, random walk, edge modeling, deep learning and hybrid.  Goyal et al. \citep{DBLP:journals/corr/GoyalF17}  and Cui \citep{cui2018survey} present the graph embedding techniques in three categories: factorization, random walk and deep learning based. 

Unlike previous studies, we employ a graph theory to learn nodes' representations. By using the dominating set theory, our proposed method creates intelligent random walks that can preserve both local and global information. 

\section{Proposed Method}\label{Proposed_Method}
\begin{figure}
	\centering
	\includegraphics[width=0.8\textwidth]{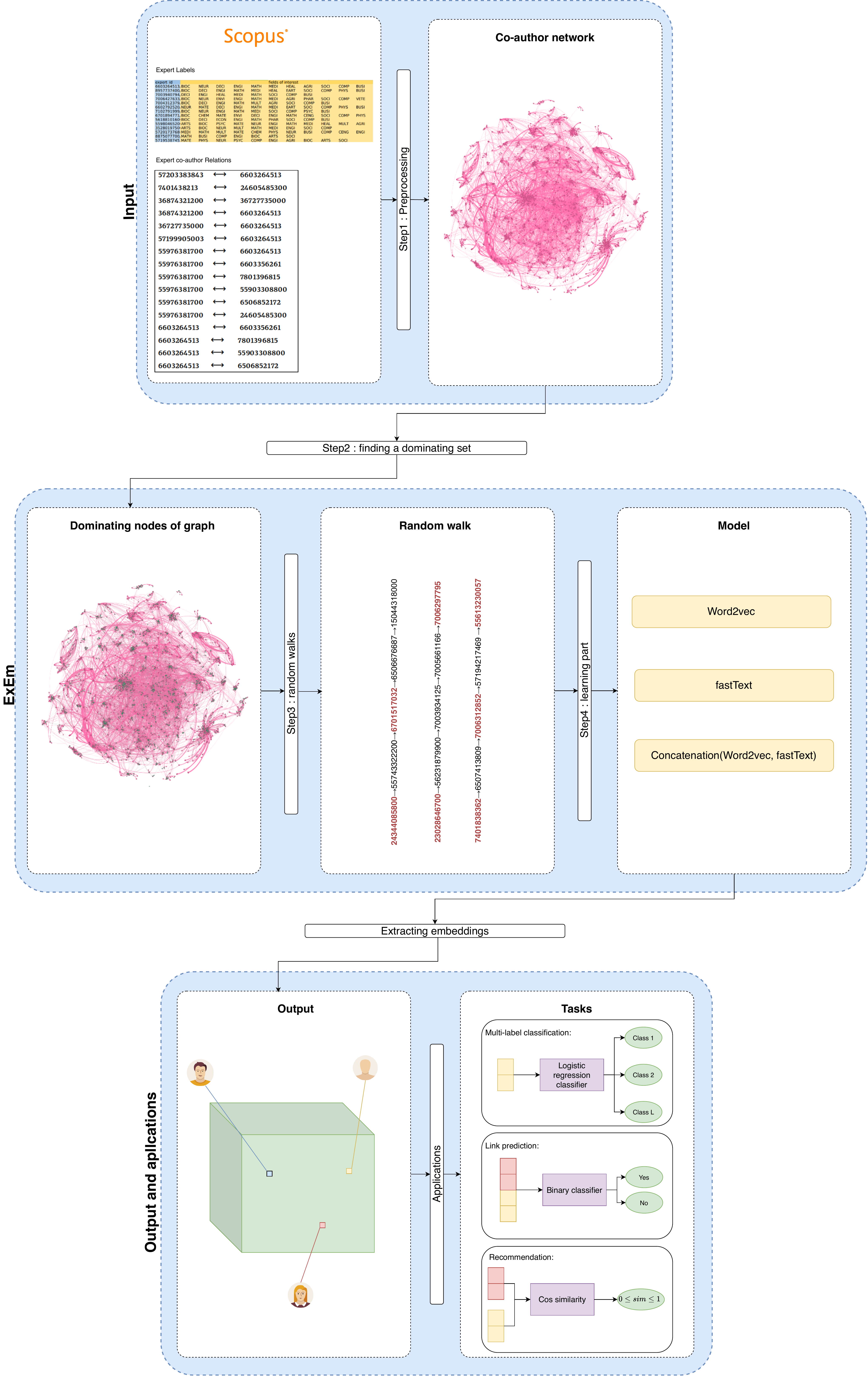}
	\caption{The overall structure of ExEm.}
	\label{diagram}
\end{figure}

The aim of our study is to incorporate the dominating set concept from graph theory to the graph embedding. We propose a new model, which is called ExEm, that is able to map a graph, our case study a  co-authorship network, to a low-dimensional vector space.  The overall structure of ExEm is shown in Figure \ref{diagram}. ExEm initially extracts the adequate dataset from  \footnote{\url {https://www.Scopus.com}}{Scopus} which is the largest abstract and citation database. The gathered dataset includes the features of expert candidates such as their subject areas, affiliations, h-index, and their co-author interactions. In the next phase, ExEm converts the extracted information into a labeled collaborative network where nodes and their labels represent authors and their subject areas, and edges show authors' co-author collaborations. Then, ExEm gets the constructed graph as input and applies the dominating set theory on it. Since dominating set acts as a backbone and governs the graph, it enables ExEm to create comprehensive and meaningful representation of a graph. To capture nodes' representations, ExEm constructs intelligent random walks that comprise of at least two dominating nodes. One dominating node should appear at the beginning of each path sampled to characterize the local neighborhoods. While, the other one reflects the global structure information of a graph. Finally, ExEm adapts a skip-gram neural network to obtain the node embeddings. To train the skip-gram model, ExEm exploits three embedding methods including Word2vec, fastText and the concatenation of these two. The embedding results can be applied to many applications such as multi-label classification, link prediction, and node recommendation, which can achieve much better performance than existing graph embedding approaches. The following subsections describe the procedures of ExEm in detail.

\subsection{Step 1: preprocessing}
Data preprocessing plays an important role in the social network analysis. The goal of preprocessing is to convert the original dataset to an acceptable format for discovering beneficial information or recognizing patterns from the social network dataset \citep{gupta2013data}. In this way, the first step of ExEm is preprocessing the dataset. So, the relations between the nodes of the dataset are considered and the graph of the dataset is constructed. Also, nodes may contain assigned elements such as attributes, labels, and tags. So, the output of this step is a graph $G = <V,L,E>$ that $V$, $L$ and $E$ demonstrate the nodes of the graph, the corresponding element values of nodes, and edges between nodes, receptively.  The first block of Figure \ref{diagram} (noted as input) shows the preprocessing task that is applied on Scopus dataset. As it is observable from this figure, ExEm transforms the extracted information from Scopus which includes experts id, their fields of interest, and their connections into a labeled collaborative graph format. Experts and their subject areas are defined as nodes and their labels, respectively.  The graph edges originate from the authors' co-author collaborations.

\subsection{Step 2: finding a dominating set} 
This step aims to find a dominating set  (DS) of the corresponding graph $G$ which was created in the previous step.  A subnet of nodes, $D$, is called a DS if every node is either in $D$ or adjacent to a node in $D$ \citep{du2012connected}. Research demonstrates that DS constructs a virtual backbone on the graph and plays an important role in monitoring and controlling the connections of nodes \citep{sun2019minimum}.  Dominating sets are able to perform various critical tasks in different networks such as the study of social influence propagation in social networks or finding high-impact optimized subsets in protein interaction networks \citep{molnar2014dominating}. 

\begin{figure}
	\centering
	\includegraphics[width=0.4\textwidth]{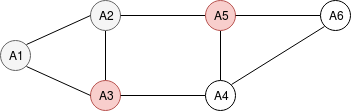}
	\caption{An example of a graph with 6 nodes and 5 edges and its dominating set.}
	\label{ex-ds}
\end{figure}

Since finding a DS is classical NP-complete decision problem, many greedy, approximation and heuristic approaches have been proposed to detect a dominating set in the given graph. Talking about the advantages and disadvantages of these techniques and investigating the best solution for constructing the DS is out of the scope of this paper. In ExEm, dominating set $D$ is produced by the algorithm $7$ in \citep{esfahanian2013connectivity} that is a simple and distributed approach.  The pseudo code of this algorithm is shown in Algorithm \ref{alg1}. Based on the algorithm, one of the nodes is randomly selected and added to the dominating set $D$. After that, this node and its neighbours are removed from the graph nodes. Then, another random node is chosen from remaining nodes and inserted into $D$. The mentioned steps are continued until there is no node in graph node set $V$.  As an example, in Figure \ref{ex-ds},  after applying the dominating set algorithm on the graph, one possible selection set of dominating nodes is $A_3$ and $A_5$.  It is obvious that all nodes in the graph are accessible by $A_3$ and $A_5$.

\begin{algorithm}
	\centering
	\caption{Finding a dominating set}
	\begin{algorithmic}\label{alg1}
		\REQUIRE {A connected non–trivial graph $G = (V,E)$}
		\STATE $D=\emptyset$
		\LOOP
		\IF {\textbf{IsEmpty(}$V-[D \cup Neighbors(D)]$\textbf{)}}
		\STATE STOP
		\ENDIF
		\STATE Select randomly a vertex $w$ $\in$ $V-[D \cup Neighbors(D)]$
		\STATE $D$ $\leftarrow$ $D$ $\cup$ $\left\{w\right\}$
		\ENDLOOP
		\RETURN $D$
	\end{algorithmic}	
\end{algorithm}

Furthermore, the first sub-block (from left to right) of second block in Figure \ref{diagram} indicates the result of applying the dominating set theory on Scopus graph. Green nodes represent dominating nodes.

Ideally a graph embedding approach should fulfill two objectives including homophily and structural role. Homophily indicates the tendency of each pair of nodes in the graph to be similar. Based on homophily, all nodes in a community should have similar embeddings. On the other hand, the structural role objective ensures that the nodes with the similar functions in the graph, should be embedded closely together. In other words, homophily emphasizes connectivity, whereas the nodes in a particular structural role  may inhabit in very different parts of a graph.  We observe that using dominating nodes allows a flexible method that can convert graph nodes into vectors obeying the above equivalences. Dominating nodes are an approximation of the whole network and administer the rest of nodes. As another way of looking, the DS allows a clustering scheme on the graph that dominating nodes operate as cluster heads from which all nodes in the cluster can be reached in one-hop. As each dominating node supervises the nodes of the community which it belongs to, it can obtain a local view of the underlying subset. Consequently, this utility promotes ExEm to properly learn that dominating nodes and their dominated nodes should share similar embeddings  because of pertaining to the same cluster or community. For example, in Figure \ref{ex-ds}, we can see that nodes $A_3$ and $A_1$ have similar neighbourhoods and are a part of a community. So, ExEm achieves the homophily target by embedding these two nodes to similar vector representations. Additionally, dominating nodes provide a backbone between communities. The connections between these backbones develop the awareness of the global graph structure. That is to say that the choice of DS for networks as the virtual backbones facilitates ExEm to accomplish the structural role objective. 
In Figure \ref{ex-ds}, nodes $A_3$ and $A_5$ are not close in terms of graph distance, but they have similar local structural roles.  ExEm coverts these nodes into similar vectors because both of them play the same roles as the heads of their communities.  The other key advantage of DS is that no global information is required to construct it. The employed algorithm for finding dominating set uses only local information to get the DS and based on the studies \citep{yang2003dominating, esfahanian2013connectivity} it is shown to be the fastest one. 

\subsection{Step 3: random walks} 
With having the dominating nodes from the previous step, we introduce our intelligent random walk strategy in this subsection. Before giving full details of the proposed random walks, we are going to describe what random walk is and why it is important in graph embedding. A random walk on the graph is defined as a random sequence of nodes where consecutive nodes are neighbors \citep{liu2016smartwalk}. Random walks can obtain the information hidden in the graph structure. 
The importance of random walks in graph embedding domain is adopted from natural language processing (nlp) after great success of word embedding models. In graph embedding, the graph properties are preserved by a set of random walk paths sampled from it \citep{cai2018comprehensive}. In other words, each random walk in the graph embedding presents other concept which is the equivalent of a sentence definition in nlp domain. That means random walk and sentence have the same responsibilities in their scopes. Additionally, the nodes of the random walk take on the role of words or vocabularies in the sentence. There are some advantages of random walk based graph embedding approaches including the acceptable level of time and space complexity \citep{pimentel2019efficient}, no need for feature engineering, and investigation of diverse parts of the same graph at the same time  by a number of sampled paths\citep{grover2016Node2vec, cai2018comprehensive, liu2016smartwalk}. Hence, many graph embedding methods have been proposed based on random walks such as DeepWalk and Node2vec where their difference comes from their sampling strategies. However, these approaches suffer from finding optimal sampling procedure. DeepWalk uses a uniform random walks which can not control over the search space. Node2vec suggests a biased random walks in which some node neighbors have a higher or lower probability of being selected in each step by two parameters. The problem is finding the best values for these parameters which determine the likelihood of observing nodes in the each random walk for every network. 

ExEm is a random walk based technique that modifies the random walk strategy used in DeepWalk and Node2vec by hiring dominating nodes. Our proposed intelligent random walks offer the flexibility in sampling nodes from a network.  The concept of this intelligence emanates from appearing two dominating nodes in the sampled paths. For each random walk, ExEm starts the path by randomly selecting one node from the dominating set found in the previous step. Then, one of the neighbors of this dominating node is chosen by chance and added to the walk. After that, walk moves to the neighbors of the last added node. The procedure of adding new nodes into the walk continues until the following two conditions are met. The main condition is the appearance of at least another dominating node in a sampled path. The other requirement for ending the process is achieving the fixed lengths $L_R$.  The second sub-block in the second block of Figure \ref{diagram} shows the examples of random paths created by ExEm from Scopus graph. In this instance, each node is presented by an expert id, red nodes indicate dominating nodes and the length of walk equals $5$. Obviously, each walk starts with a dominating node and the second dominating node can be visited in different places of a walk except the second position. The explanation is that based on Algorithm \ref{alg1}, in  the process of finding dominating nodes,  we remove a node and its neighbours after adding this node into the dominating set. So, there are not any dominating nodes in $one-$hop of this node and other dominating nodes.  Note that it is possible to see more than two dominating nodes in each path like the third random walk in the figure.

With the presence of these two dominating nodes, we see a reduction in runtime of creating random walk process. There are two aspects of how using dominating nodes decreases runtime.  Starting the random walk with a node from dominating set instead of the graph set nodes reduces the size of the search space from $|V|$ to $|D|$ where $|V|$ and $|D|$ show the sizes of graph nodes and DS, respectively. Also, after appending the first dominating node into the sample, we should add $L_R-1$ nodes to our walks to reach the maximum length $L_R$ . On the other hand, based on our strategy, we should have one dominating node in the rest length $L_R-1$ to fulfill the condition and finish adding nodes to the random walk. Note that all nodes have a uniform probability of being chosen in a walk and also the probability of selected node being a dominating is $\frac{|D|}{|V|}$. Therefore, the probability of the absence of a dominating node in the length $L_R-1$ is equal to $L_R-1$ trials and all of them are non-dominating node. The probability of each node being dominating directly affects this paradigm and it is a large number according to straight forward computation of Algorithm \ref{alg1}. Based on the explanation, it is not necessary to investigate the existence of the second dominating node in each random walk. In this way, the execution time significantly reduces.

Additionally, we observe that by the help of dominating nodes, ExEm can convert graph nodes into low-dimensional vectors obeying the homophily and structural role equivalences. The first dominating nodes ensures that ExEm selects a node within this dominating node community; so ExEm learns the node representations with respect to homophily and embeds nodes of a community into similar vectors. What it means that due to the first dominating node in the random walk, the   local neighborhoods are depicted accurately.  Moreover, this condition increases the probability of repeating nodes in the sampled neighborhoods plenty of times because each node has at least one neighbor from the dominating set. On the other hand, there are two reasons why ExEm selects the second dominating node in its random walks. The first philosophy behind it is that our sampled paths observe nodes which are far from starting node and belong to the other clusters. The algorithm of finding the dominating set proves this outlook. Since after inserting a node into DS, this node and its neighbors are removed from the node set in each step of this algorithm, there is no dominating node in $one-$hop of each dominating node. So, the second dominating node assists ExEm to preserve the global structural information of the graph. The next wisdom for the existence of the second dominating node is that dominating nodes are the heads of their communities and have the same roles to play. This allows ExEm to perceive the nodes with the same roles in each sampled path and understand that these node should be embedded closer and this is what the structural role objective emphasizes on.  So, the mentioned details confirm why using dominating set theory in creating random walks enables a flexible method that can well characterize the local and global network structure in generating node embeddings.  

Moreover, ExEm can adapt to the graph topology changes and present its dynamic characteristics. When a node is added or dropped from the graph, only its neighbors will be notified. For instance, for a new coming node, if a dominating node is within its neighborhood, ExEm constructs walks that start from its neighbor dominating node and adds them to the corpus; otherwise, it is itself considered as a dominating node and walks start from it. This demonstrates that it is just necessary to add new random walks from the changed part instead of the whole graph.

\subsection{Step 4: learning part} 
\begin{figure}
	\centering
	\includegraphics[width=0.6\textwidth]{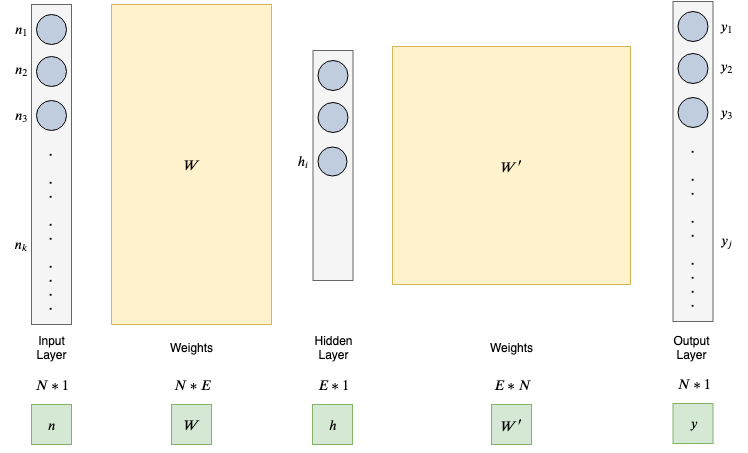}
	\caption{Skip-gram architecture.}
	\label{skip}
\end{figure}

The only required input of this step is a corpus which is created from the intelligent random walks of previous step. As mentioned before, in random walk models, node and random walk are regarded as a word and sentence, respectively. Hence, the neighborhood of a node can be observed as the co-occurrence of words in the sentence. Furthermore, there are many deep learning based approaches that can map the word co-occurrences into vector-space model. One of the most simplest and efficient techniques is skip-gram model \citep{mikolov2013distributed}. The aim of skip-gram is predicting the words surrounding a target word. The same effort can be performed in graph embedding. Accordingly, in graph embedding, the skip-gram counts the number of times node $j$ appears within a certain window of $w$. For instance, in the random walks ``$n_1 \, n_2 \, n_3 \, n_4 \, n_5$", the skip-gram gets node ``$n_3$" as input, and predicts the output ``$n_1$", ``$n_2$", ``$n_4$", and ``$n_5$", assuming $w$ is $5$.  Ski-gram architecture is a feed-forward network that is the simplest deep learning model for node representations. As shown in Figure \ref{skip} this model views a graph as a bag of nodes. For a node $n_i$, it captures a $E$-dimensional vector $y_i$ using an embedding model such as Word2vec \citep{mikolov2013efficient} or fastText \citep{joulin2016fasttext}.  Word2vec learns to convert the nodes that appear in similar random walks to similar vector representations. While, fastText takes the advantage of a bag of $n-$grams as extra features to obtain local node order information.

Considering the above explanations, in this step of ExEm, random walks from the previous step are injected as corpus into the input of the skip-gram network. ExEm exploits three embedding methods including Word2vec, fastText and the concatenation of these two to extract embeddings, as presented in the third sub-block in the second block of Figure \ref{diagram}.  There are two important points in this step that should be noted. The first one is that there are at least three common ways to combine embedding vectors and create a single vector including: summing, averaging and concatenating \citep{Damoulas2009}. In this study, we consider the concatenation of two embeddings as the basic combination approach and for further investigation, we test the summing and averaging of Word2vec and fastText embeddings in the evaluation results.  The second subject is that in skip-gram model the context window has an important effect on the resulting vector representations. The context window defines which neighbours are kept in mind when computing the vector representations \citep{lison2017redefining}. 
Therefore, having at least two dominating nodes in the context window ensures that ExEm properly understands the local and global graph information and respects for homophily and structural role objectives. Due to this procedure is in the manner of sampling each node $n_j$ with a probability that relies on the distance $|j-i|$ to the focus node $n_i$, as proved by \citep{lison2017redefining}:

\begin{equation}
	p \left( n_{i} \mid n_{j} \right) = \sum_{w=1}^{W} p \left( n_{i} \mid n_{j}, w \right)p(w) \\
	= \frac{1}{w}(w-|j-i|+1)
\end{equation} \label{pro_second_ds}

where $w$ is the real window size from $1$ to $W$. For example, with the window size $5$, the second dominating node at the position $3$ will be sampled with the probability of $\frac{3}{5}$ in Word2vec \citep{lison2017redefining}. In other words, skip-gram model maximizes the co-occurrence probability among dominating nodes that exist within a window $w$ \citep{cai2018comprehensive}

\subsection{Output and applications}\label{case_std_task}
The result of learning step is the semantic embeddings of graph nodes. As the first sub-block of the third block in Figure \ref{diagram} shows, the output of ExEm on Scopus graph is an expert embedding vector. In this way, experts of the same subject area are embedded into the similar vectors. The learned ExEm representations perform some simple algebraic operations on expert embeddings. For example, if we denote the vectors for two experts $i$ and $j$  with subject areas nlp and ml (machine learning), and ml as $E_{i(nlp,ml)}$ and $E_{k(nlp)}$, respectively, we observe that $E_{i(nlp,ml)} - E_{j(ml)} = E_{k(nlp)}$. As another instance, $E_{x(nlp)} + E_{y(bio)} = E_{z(nlp,bio)}$ results in an expert embedding that focuses on nlp approaches in bioinformatics research.

The last part is providing evaluations on ExEm with regard to its capability on real-world applications. The reason for these experiments is that a good graph embedding method should be able to effectively perform on the tasks including multi-label classification, link prediction and recommendation using the obtained representations. In the next paragraphs, we enumerate the characteristics of these tasks.

\textbf{Multi-label classification}: One of the tasks increasingly used by modern applications is multi-label classification. In this task, it is assumed that each node in the graph is associated with one or more labels from a limited set $L$ \citep{tsoumakas2007multi}. To conduct multi-label classification task, we have a model that is trained with a portion of nodes and all their labels. Then, the model gets the node representations to predict the labels for the rest of nodes. As presented in the first row of the second sub-block related to third block in Figure \ref{diagram}, a classifier like Logistic Regression is applied on a certain fraction of the expert embeddings whose subject areas are known. Then, the model predicts the subject areas for the remaining experts. That means that with help of expert embeddings and multi-label classification task we can anticipate the subject areas of experts whom no specific information is available, and only their co-author connections with other experts are provided.


\textbf{Link prediction}: Because the low-dimensional vectors of nodes encode rich information about the network structure, 
we can analyze the efficacy of predictive capacity of various embedding models through a link prediction task \citep{cai2018comprehensive}. To perform the link prediction, we arbitrarily conceal a fraction of the existing links between nodes and  
our desire is to predict these missing edges by using the node embeddings \citep{wang2016structural, grover2016Node2vec}. 
As investigated in the study \citep{chen2018link}, the link prediction can be addressed as a binary classification problem. In this case, a pair of nodes is labeled as positive if a link exists between the nodes. On the other hand, if there is no link between the node pair, then the label of the paired node is negative. As shown in the second row of the second sub-block related to the third block in Figure \ref{diagram}, two node embeddings are fed into the binary classifier. The output of the classifier is ``\textit{yes}'' if there is a link connecting the nodes, otherwise, the result is ``\textit{no}''.  Thus, better results can be retrieved in the link prediction task by using a graph embedding technique that learns a deep representation of the nodes on the network.

Also, we can explore the link prediction potential usage in the expert recommendation system. For this purpose, the classifier accepts two expert embeddings as inputs and anticipates that these two experts can whether be co-authors or not. If they have similar expert embeddings, which show their expertise is closed, then the classification result is ``\textit{yes}''.

\begin{figure}
	\begin{center}
		\centering
		\includegraphics[width=1\textwidth]{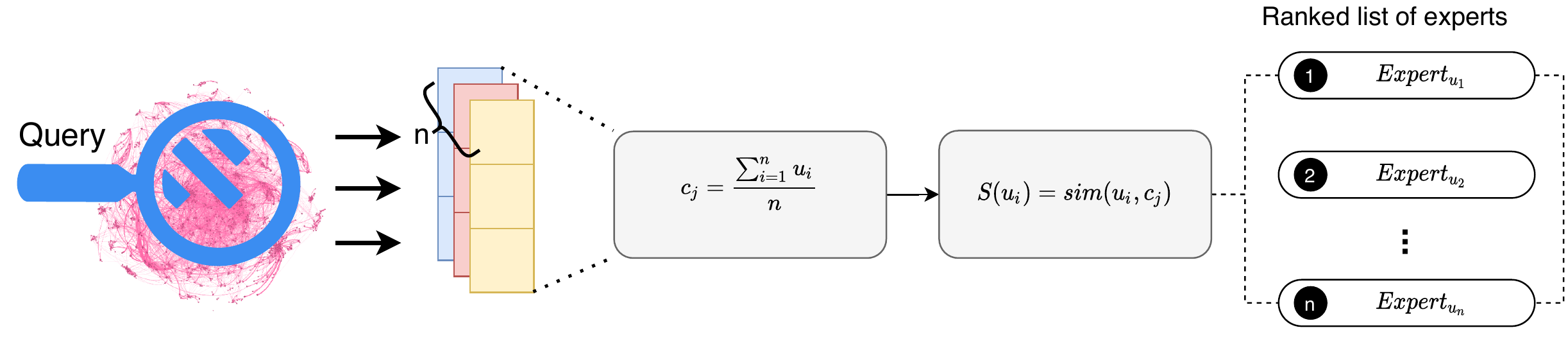}
		\caption{The process of our proposed method for computing experts' scores in an expert recommendation system.}
		\label{case-study}
	\end{center}
\end{figure}

\textbf{Recommendation}\label{case_std}: Graph embedding approaches have demonstrated to be beneficial for the node recommendation that is the task of recommending top nodes of interest to a given query according to certain specifications \citep{cai2018comprehensive}. To extend the graph embedding algorithms specifically ExEm into the recommendation task, we need a strategy for computing nodes' scores and ranking nodes by using the generated vector representations of the nodes. In the following paragraphs, we introduce how ExEm and other graph embeddings can leverage the embeddings for the expert recommendation task by proposing a novel scheme. It should be noted that the types of recommended nodes are miscellaneous and the proposed procedure can be applied to them with a few changes. In this paper, the recommendation items are experts whose research interests and expertise are most similar to a given topic. Clearly, an expert recommendation system takes a user's query in the term of input and then provides a list of experts sorted by the degree of their relevant expertise with the given query \citep{NIKZADKHASMAKHI2019126}. Figure \ref{case-study} indicates our proposed method to make recommendation experts based on expert embeddings. The user's query, that is a topic, is injected into the input of the recommendation system. Then, experts whose subject areas include this topic are extracted to make a cluster. Note that we can predict the subject areas of experts with unknown labels through multi-label classification task by using experts' low-dimensional vectors.  After constructing the community, the center of this cluster is found by taking the average of all the expert embedding vectors in the group. Finally, the similarity measure functions such as Euclidean, Cosine, and Manhattan can be employed to calculate the distance between each expert and the centroid. This similarity is considered as an expert's score.


\section{Data Description}\label{data_dis}
\begin{figure}
	\begin{center}
		\includegraphics[width=0.7\textwidth]{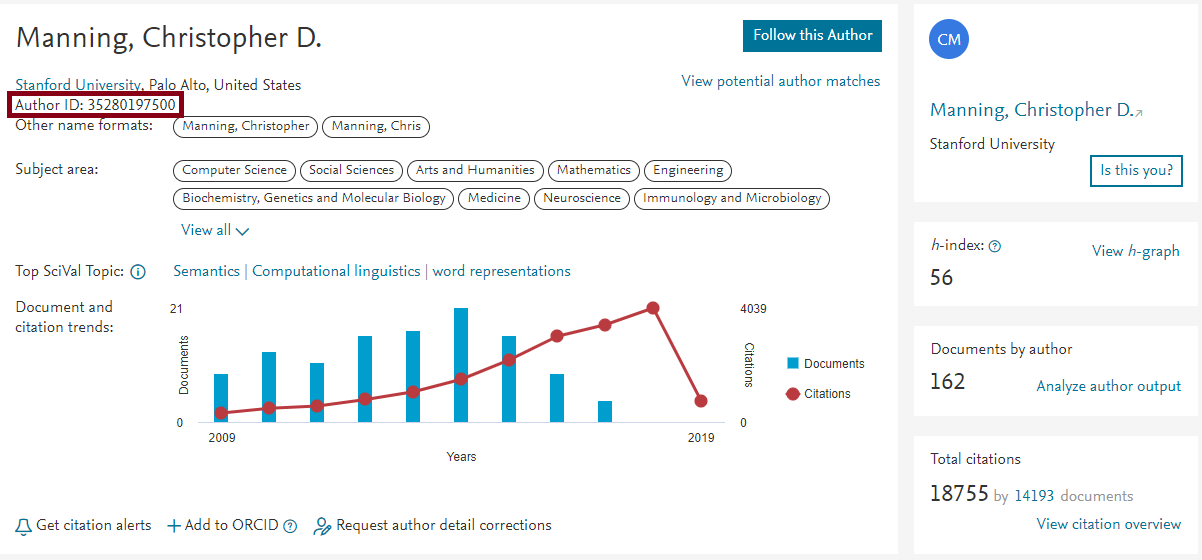}
		\caption{An author profile in Scopus.}
		\label{profile}       
	\end{center}
\end{figure}

As mentioned before, we also gathered a collaborative network in this paper. We figured out \textit{Scopus} is an adequate source that consists of a wide number of authors and their articles from scientific areas. 
Authors with publications indexed in Scopus have their own profiles and a unique Scopus author identifier. Figure \ref{profile} shows an example of an author's profile in Scopus.  Different types of information can be extracted from the authors' profiles. This information includes the content and non-content features of authors such as their published articles, subject areas, affiliations, h-index, co-authors and number of citations of each paper. We use a part of this data and build a co-author network. In this network, we call authors as experts. Therefore, experts' ids and their subject areas are presented as graph nodes and their labels. Also, the experts' co-author collaborations form the graph edges.

To create this labeled collaborative network, there are several motivations that are enumerated in the following. Recently, there has been an increasing interest in graph embedding techniques. The obtained representations from graph embedding methods are evaluated on learning tasks such as multi-label classification. Karate, BlogCatalog, Wikipedia, and Protein\textendash Protein Interactions(PPI) are the most used labeled datasets to estimate the efficiency of a proposed graph embedding approach on multi-label classification task. These labeled datasets are types of social networks and biology networks. 
Lack of a labeled collaborative network is felt in testing graph embedding approaches. 
Moreover, there is a demand on a labeled collaborative network dataset for the usage in supervised machine learning methods in expert finding system or detecting communities of experts in collaborative networks. 
In summary, the usage of our collected dataset can be listed as: multi-label classification, link prediction, recommendation, community detection,  and expert finding tasks.

To collect data, we initially selected $20$ experts from the \href{https://aminer.org/lab-datasets/expertfinding/}{\textit{Arnetminer}} expert list related to ``Information Extraction" topic and obtained these experts' information from Scopus. Then, we extended the extraction of information related to the co-authors with a $two-$hop expansion. It means that we gathered the information of co-authors of these experts and the co-authors of these co-authors in the next steps.

\begin{figure}
	\centering
	\subfloat[Graph visualization of Scopus.]{
		\includegraphics[width=0.4\textwidth]{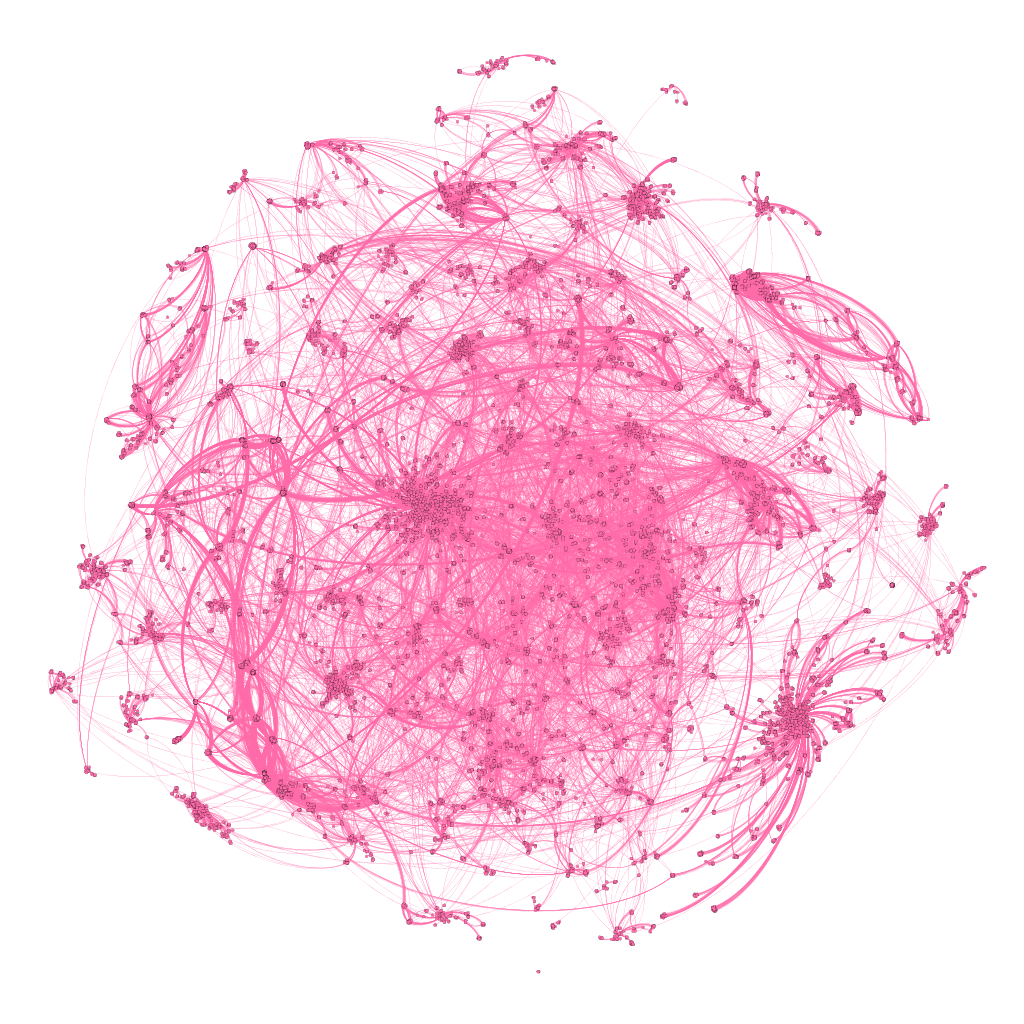}
		\label{OnlyGraph}}
	\subfloat[Dominating nodes of Scopus.]{
		\includegraphics[width=0.4\textwidth]{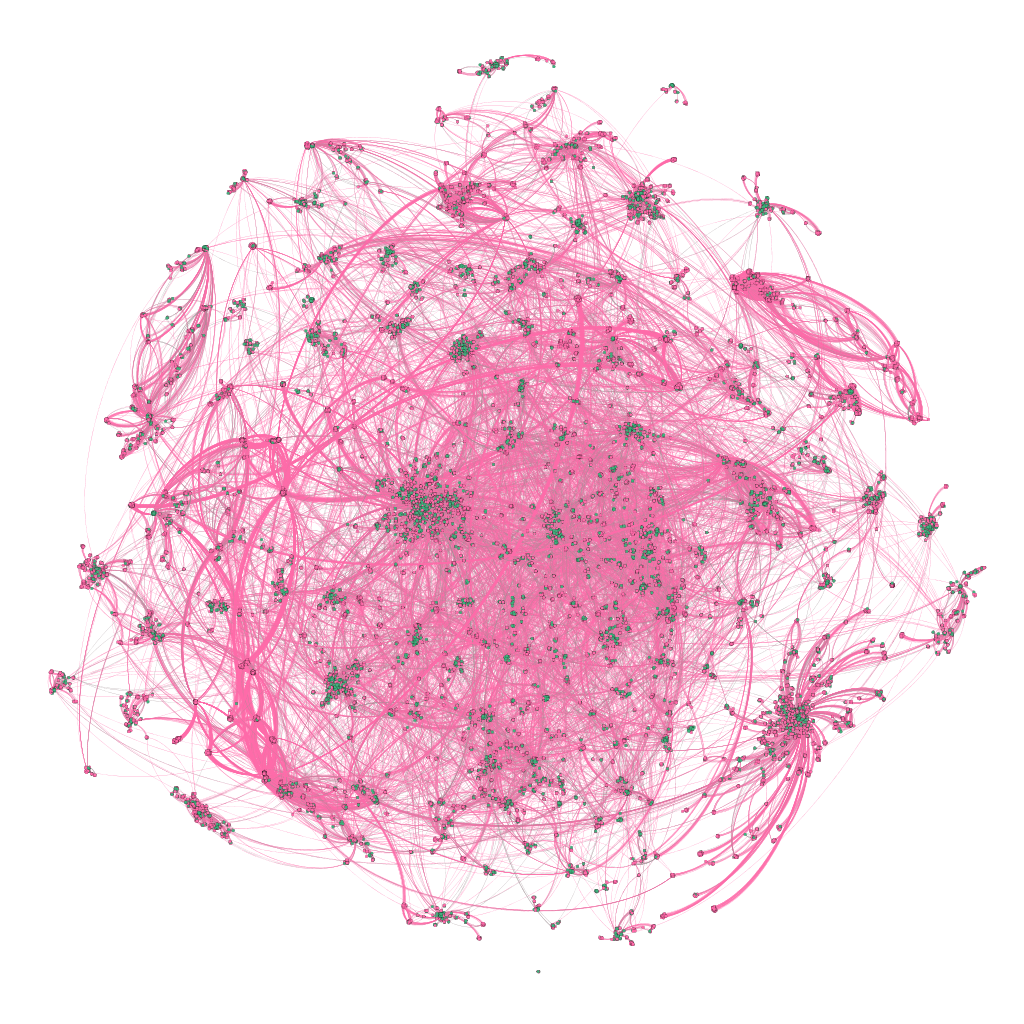}
		\label{ds_Scopus}}
	\qquad
	\subfloat[Visualization of experts: larger numbers in size denote higher degree of author.]{
		\includegraphics[width=0.4\textwidth]{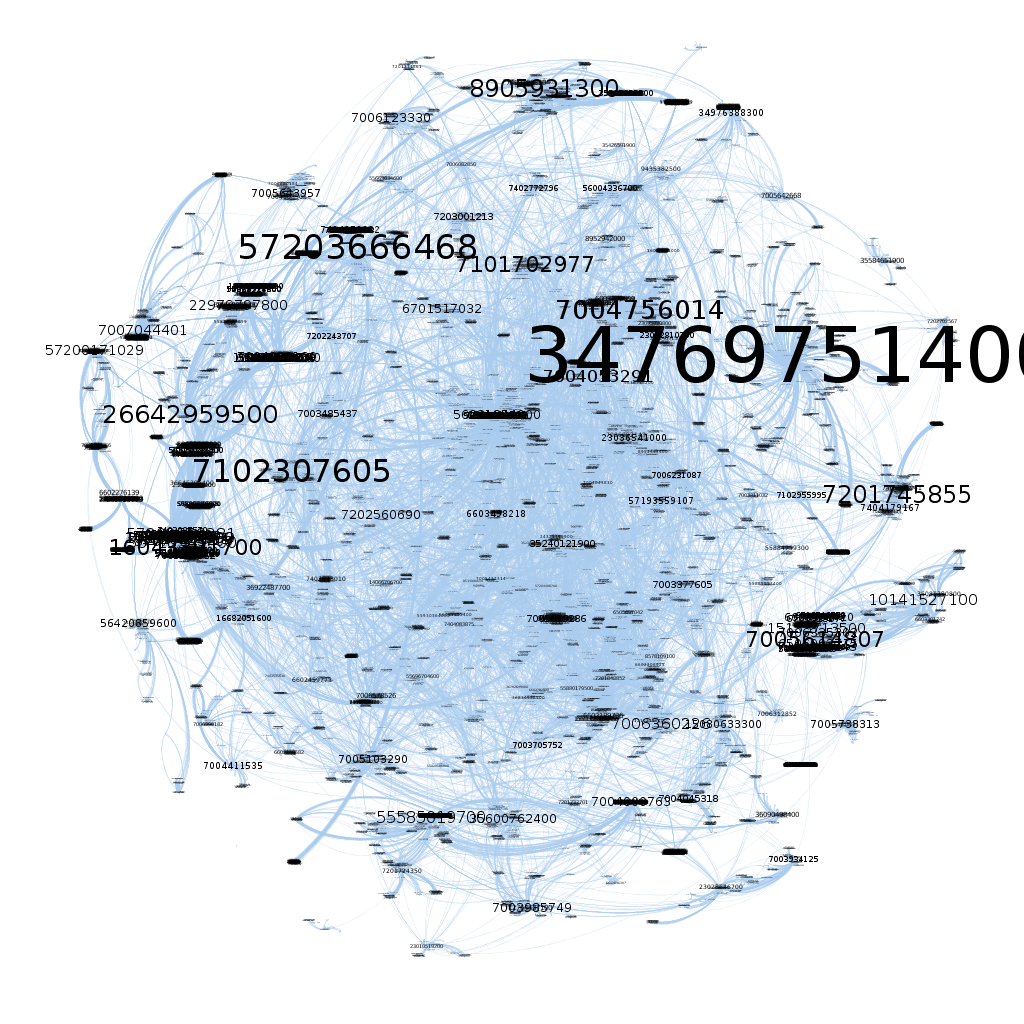}
		\label{GraphbigestAuthors} }
	\subfloat[Visualization of the detected communities.]{
		\includegraphics[width=0.4\textwidth]{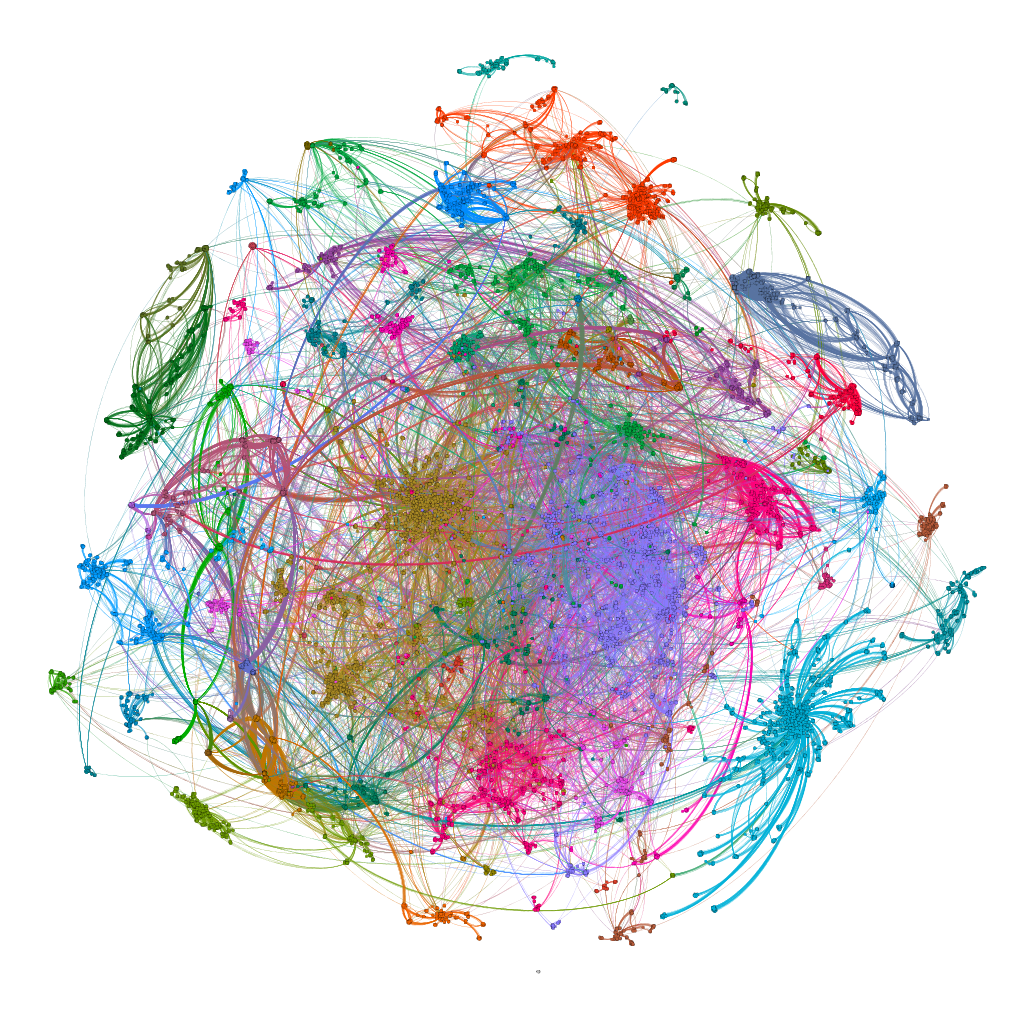}
		\label{GraphCommunity}}
	
	\caption{Various visualizations of Scopus graph.} 
	\label{fig:gra}
\end{figure}

To provide a clear understanding of our constructed network, we have shown structural information of this graph as diagrams using \href{https://gephi.org/}{\textit{Gephi}} which is an open-source network analysis and visualization software \citep{bastian2009gephi}. Figure \ref{OnlyGraph} presents the visualization of the created collaborative network from gathered data. Dominating experts of this network are highlighted in Figure \ref{ds_Scopus}. Moreover, Figure \ref{GraphbigestAuthors} is the visualization of experts by this overview that the larger numbers in size of the expert' identifier denotes the higher degree of the expert. Based on this representation, the expert  with id $34769751400$ is the one with the highest degree, $2147$, in the graph. On the other hand, Figure \ref{GraphCommunity} displays the communities detected by applying the proposed method in the study \citep{blondel2008fast} on Scopus graph; nodes are colored according to their communities.  The value of modularity of our constructed graph is $0.912$ that exposes Scopus graph has dense connections between the experts within communities and sparse connections between experts in different communities.  Also, the average clustering coefficient is $0.889$ that shows the tendency of experts to cluster together.  Finally, Table \ref{label_dist} shows how many experts belong to each label. It can be observed that Scopus graph covers experts from different scientific areas and also the most number of experts have label "COMP".  It should be noted that labels with a higher percentage of $5 \%$ are listed in this table.  Summary of Scopus dataset is demonstrated in Table \ref{dataset_statistics}. 

\begin{table}
	\caption{Percentage of authors related to each label.}
	\label{label_dist}
	\centering
	\begin{tabu}{p{1cm}|p{2cm}}
		\hline
		Label & Percentage\\
		\hline
		COMP & $12.81 \%$\\
		\hline
		ENGI & $10.28 \%$\\
		\hline
		MEDI & $7.49 \%$\\
		\hline
		BIOC & $7.08 \%$\\
		\hline		
		SOCI & $6.16 \%$\\ 
		\hline
		DECI & $5.74 \%$\\
		\hline
		ARTS & $5.71 \%$\\
		\hline
		NEUR & $5.22 \%$\\
		\hline
	\end{tabu}
\end{table}

\begin{table}
	\caption{Summary of Scopus dataset.}
	\label{dataset_statistics}
	\centering
	\begin{tabu}{p{5cm}|p{1cm}}
		\hline
		Number of nodes &$27473$\\
		\hline
		Number of edges &$285231$\\
		\hline
		Average degree&$20.7645$\\
		\hline
		Maximum degree&$2147$\\
		\hline		
		Average clustering coefficient& $0.889$\\ 
		\hline
		Number of triangles& $4582111$\\
		\hline
		Modularity&$0.912$\\
		\hline
		Modularity with resolution&$0.912$\\
		\hline
		Number of Communities&$49$\\
		\hline
		Number of labels&$27$\\
		\hline
	\end{tabu}
\end{table}

\section{Experimental Evaluation} \label{sec:experimentalEvaluation}
In the present section, we will provide an overview of the datasets on which the ExEm is applied. Next, 
we will introduce four baseline algorithms to compare ExEm against them. Then, we are going to describe the used parameter settings. Finally, we will specify the metrics hired to evaluate our proposed algorithm.

\subsection{Dataset}
In the succeeding paragraphs, we are going to characterize the datasets on which our experiments were conducted.

\textbf{BlogCatalog} \citep{zafarani2009social}: This is a social blog directory where nodes demonstrate the bloggers and edges show the friendship connection among the bloggers. Each blogger is labelled by at least one category that represents the blogger's interests.

\textbf{Protein–Protein interactions(PPI)} \citep{breitkreutz2007biogrid}: This is a biological network. In this graph, nodes are proteins and edges indicate the pairwise physical interactions between proteins in humans. The labels of nodes are obtained from the protein-coding gene sets.  

\textbf{Wikipedia} \citep{mahoney2011large}: This is a network of co-occurrence words related to Wikipedia's articles. The node labels indicate the Part-of-Speech (POS) tags assigned to the node. 

\textbf{arXiv(Astro-PH)}\citep{leskovec2014snap}: This is a collaborative network that is constructed from the collaborations between authors' papers submitted to the e-print arXiv and Astro Physics category. On the other hand, the nodes of this graph determine authors and edges express the co-authored relationships between authors.

We also evaluated the performance of our algorithm on Scopus dataset. The descriptions of all datasets are summarized in Table \ref{tb_dataset}. The column task shows the tasks hired to evaluate methods on that specific dataset. The details of tasks were represented in subsection \ref{case_std_task}.

\begin{table}[h]
	\centering
	\caption{Summary descriptions of datasets.}
	\label{tb_dataset}
	\begin{tabu}{l| c| c |c| c|c|c}
		\hline
		Name  &  $|V|$ & $|E|$ & Labels & \multicolumn{3}{c}{Task}\\
		\cline{5-7}
		& & & & Multi-label classification & Link prediction & Recommendation \\
		\hline
		BlogCatalog	& 10,312 & 333,983	& 39 & \checkmark & \checkmark & -\\
		PPI & 3,890 & 76,584 & 50 & \checkmark & \checkmark & -\\
		Wikipedia & 4,777 & 184,812 & 40 & \checkmark & - & - \\
		arXiv(Astro-PH) & 18,772 & 396,160 & - & - & \checkmark &-\\
		Scopus & 27,473 & 285,231 & 27 & \checkmark & \checkmark & \checkmark\\
		\hline
	\end{tabu}
\end{table}

\subsection{Model Variations}
We have experimented with several variants of the ExEm model.

\textbf{ExEm$_{ft}$}: It is a version of ExEm that engages fastText method to learn the node representation.

\textbf{ExEm$_{w2v}$}: This one is another form of ExEm that allows to create node representations by using Word2vec approach.

\textbf{ExEm$_{com}$}: It is defined as the concatenation of the node embeddings learned by ExEm$_{ft}$ and ExEm$_{w2v}$.

\textbf{ExEm$_{sum}$}: It is the other model of ExEm that creates node embeddings by computing the addition of ExEm$_{ft}$ and ExEm$_{w2v}$ vectors.

\textbf{ExEm$_{avg}$}: It applies average function to ExEm$_{ft}$ and ExEm$_{w2v}$ representations to obtain node embeddings.

\subsection{Baseline algorithms}
To approve the performance of ExEm, we will compare it against the following baselines. Among them, DeepWalk and Node2vec are deep learning graph embedding methods with random walks that capture the neighbourhood pattern of the graph through sampled paths on it \citep{cai2018comprehensive}. While, SDNE is a deep learning based graph embedding approach that applies auto-encoders on the whole graph. Finally, Line is an edge modeling based method that minimizes an objective function to preserve first- and second-order proximities.

\textbf{DeepWalk}\citep{perozzi2014deepwalk}: represents a graph as a set of simple random walks starting on each node, Then these random walks are trained using the skip-gram algorithm to  create node embeddings \citep{pimentel2019efficient}.

\textbf{Node2vec}\citep{grover2016Node2vec}: is the extended version of DeepWalk with a more elaborate random walk. Node2vec introduces a biased-random walk using the breadth-first and depth-first search techniques. Node2vec governs the search space through two pre-assigned parameters $p$ and $q$ .

\textbf{SDNE}\citep{wang2016structural} uses two auto-encoders on the whole graph to learn representation. For each node, auto-encoders are structured to take the second-order proximity as inputs and are trained to reconstruct the neighborhood structure of that node. SDNE connects two auto-encoders through a loss function, $L_{1st}$, that preserves the first-order proximity \citep{zhang2018network}. The auto-encoder loss function shown by $L_{2nd}$ and $L_{1st}$ are combined linearly to minimize the total loss of the network given by

\begin{equation}\label{SDNE}
	L =  L_{2nd} + \alpha L_{1st} + \nu L_{reg}
\end{equation}

here $L_{reg}$ represents a regularization term. 

\textbf{Line} \citep{tang2015line} is an edge modeling based method that optimizes an edge reconstruction. Three different models of Line are proposed: Line(1st), Line(2nd) and Line(1st+2nd). The objective functions of Line(1st) and Line(2nd) are designed to preserve the first-order and second-order proximities, respectively. While, Line(1st+2nd) minimizes the differences between the first- and second-order proximities. We use Line(1st+2nd) for comparison, as the original study states that it outperforms all other methods of Line.  Note that we refer to Line(1st+2nd) as Line in the whole paper .

\subsection{Parameter settings}
We optimized the optimizer with Stochastic gradient descent (SGD) and performed SGD parameters similar to the method proposed by \citep{grover2016Node2vec}. Also, for all embedding methods, we used the same parameters that are reported in \citep{grover2016Node2vec}: number of walks per node $K$: $10$;  length of random walks $L_R$: $80$; node vector dimension $E_d$: $128$ (ExEm$_{com}$: $256$); context window size $w$: $10$. Further, for Node2vec, we selected the best values of parameters $p$ and $q$ from $[0.25, 0.5, 1, 2, 4]$ as proposed in \citep{grover2016Node2vec}. For SDNE, we optimized the parameters as suggested in \citep{wang2016structural}: the architecture with $[10300, 1000, 128]$ nodes on each layer, $\alpha = 0.2$, $\beta = 10$ and $\gamma \in [1e-4, 1e-5]$. For Line, we set the numbers of iterations and negative samples to $50$ and $5$, respectively.

\subsection{Evaluation Metrics}
To assess the quality of ExEm  on node classification, we use Micro-F1 and Macro-F1 scores as our metrics. For link prediction , we use Area Under Curve (AUC) score. Finally,  Normalized Discounted Cumulative Gain (nDCG) is used to evaluate the performance of ExEm over recommendation task.  These metrics are defined as follows

F1 score can be explained as a weighted average of the precision and recall. The formula of the F1 score is presented in equation \ref{eq:f1}.

\begin{equation}\label{eq:f1}
	F1 = 2\times\frac{Pr\times Re}{Pr + Re}
\end{equation}

where $Pr$ and $Re$ denote precision and recall, accordingly. 

\textbf{Micro-F1} calculates the F1 score of the accumulated contributions of all labels. In the other words, this score highlights the common labels in the dataset by considering the equal importance for each instance. Equation \ref{eq:microf1} represents the definition of Micro-F1.

\begin{equation}\label{eq:microf1}
	Micro-F1 = 2\times\frac{ microPr\times microRe}{ microPr + microRe}
\end{equation}

here $microPr$ and $microRe$ are defined by equations \ref{eq:micropr} and \ref{eq:microre}, respectively.

\begin{equation}\label{eq:micropr}
	microPr = \frac{\sum\limits_{l\in L} TP_l}{\sum\limits_{l\in L}{(TP_l+FP_l)}}
\end{equation}

\begin{equation}\label{eq:microre}
	microRe = \frac{\sum\limits_{l\in L} TP_l}{\sum\limits_{l\in L}{(TP_l+FN_l)}}
\end{equation}

where $TP_l$ and $FN_l$ present the number of true positives and false negatives within samples which are assigned to the label $l$.

\textbf{Macro-F1} is interpreted as the mean of label-wise F1 scores. This score equally treats all labels. The low value of Macro-F1 for a model shows
that the model performs well on the common labels while it has poor performance on the rare labels. Macro-F1 is calculated as following:

\begin{equation}\label{eq:macro}
	Macro-F1 = \frac{\sum\limits_{l\in L} F1(l)}{L}
\end{equation}

where $F1(l)$ denotes the F1 score for label $l$.

\textbf{AUC} score is the most common evaluation metric to evaluate the accuracy of the prediction in the link prediction task.   AUC value reflects the probability that a randomly chosen existing link is positioned to the right of a randomly chosen non-existent link. The larger AUC score is the higher the probability that there is a connection between node $u$ and node $v$ for the pair of nodes $(u,v)$ \citep{chen2018link}. AUC is defined as

\begin{equation}\label{eq:auc}
	AUC=\frac{{n}_{1}+0.5{n}_{2}}{n}
\end{equation}

here $n, n_1$ and $n_2$ illustrate samples, samples which have a higher score for existing links, and samples have resulted in the same scores, respectively \citep{ahmad2020missing}.

\textbf{nDCG} is a ranking measurement that evaluates the gold standard ranked list of experts against the ranked list outputs from recommendation task. The more the correlation between these two ranked lists yield the higher value of nDCG. The DCG for $k$ recommendations (DCG$@k$) sums the true scores ranked in the order induced by the predicted scores, meanwhile adding a logarithmic discount. DCG$@k$ is given by

\begin{equation}\label{eq:dcg}
	DCG@k=e_{rel_i}+\sum_{i=2}^{k}\frac{e_{rel_i}}{log_2(i-1+1)}=e_{rel_i}+\sum_{i=2}^{k}\frac{e_{rel_i}}{log_2(i)}
\end{equation}

where $e_{rel_i}$ is the true relevance of the recommendation at position $i$ for the current expert $e$. Then we can obtain nDCG$@k$ as follow:

\begin{equation}\label{eq:ndcg}
	nDCG@k=\frac{DCG@k}{IDCG@k}
\end{equation}

here IDCG is the DCG of ideal order.

\section{Evaluation Results}\label{evaluation_re}
In the following paragraphs, firstly, we will evaluate and compare ExEm with other embedding methods on the three tasks presented before. For each task, we are going to present results by varying the size of the training set and, then, we will examine the effect of number of embedding dimensions on the performance. Finally, in the last subsection, we will study the parameter sensitivity of ExEm measured by the classification performance.

\subsection{Multi-label classification}\label{Multi-label_classification_sec}
Multi-label classification is one of the tasks for evaluating the performance of a graph embedding approach. A good node embedding method can give  the graph embeddings as an input and predicts the node labels. So, we valuated ExEm accomplishments under the multi-label classification task. Firstly, we captured the node embeddings of the input graph for each algorithm. The dimensions of node embedding are $256$ and $128$ for ExEm$_{com}$ and others, respectively. Then,  we randomly selected a portion ( $10 \%$ to $90 \%$) of nodes along with their labels as training data to analyze the achievements on the remaining nodes. We trained a one-vs-rest Logistic Regression classifier which was implemented by LibLinear \citep{fan2008liblinear}. For the purpose of ensuring a fair comparison, we repeated the above procedure $10$ times and reported the results in terms of average Micro-F1 and average Macro-F1. In the paragraphs that follow, firstly, we are going to present the obtained results for each dataset, then we will show the effect of number of embedding dimensions on the performance of classification task. 

\subsubsection{Results}
\begin{figure}
	\centering
	\includegraphics[width=0.9\textwidth]{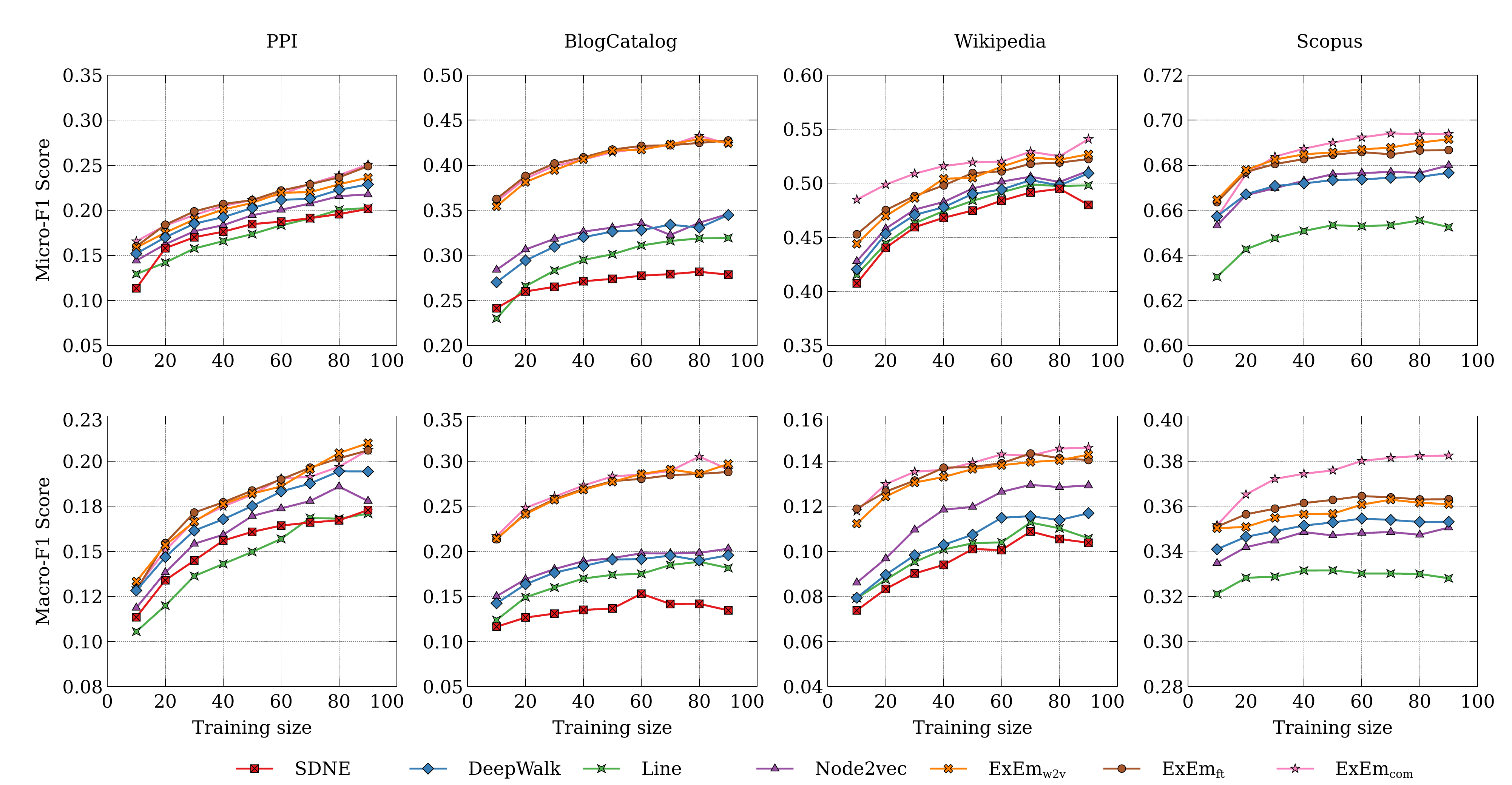}
	\caption{Micro-F1 and Macro-F1 scores on node classification task for different datasets over the diverse train ratio (dimension of ExEm$_{com}$ is $256$)} 
	\label{multi-label-calssification-re}
\end{figure}

Figure \ref{multi-label-calssification-re} shows the results of the classification task based on Micro-F1 and Macro-F1 scores for different approaches under PPI, BlogCatalog, Wikipedia and Scopus datasets. From the results, we have the following observations and analysis based on each dataset:

--\textbf{{PPI dataset}}: It is evident that various versions of ExEm gain the highest Micro-F1 and Macro-F1 scores under PPI dataset.
Given $10 \%$ of nodes as training data, as an example, ExEm outperforms DeepWalk, Node2vec, Line and SDNE on Micro-F1 by $8.94 \%$, $14.84 \%$, $28.07 \%$ and $45.90 \%$, respectively.  Also, ExEm achieves $3.89 \%$, $12.29 \%$, $26.35 \%$ and $17.44 \%$ improvements in terms of Macro-F1 over DeepWalk, Node2vec, Line and SDNE, individually. Both DeepWalk and Node2vec that are based on random walks perform better than  Line and SDNE which use first- and second-order proximities, and auto-encoders, accordingly. Additionally, the results demonstrate that the learned node embeddings of DeepWalk  can better generalize to the classification task on PPI dataset than Node2vec, since appropriate values are not assigned to Node2vec's parameters. Also, we find SDNE the winner of the competition against Line. 

--\textbf{{BlogCatalog dataset}}: We  have observed  that  using dominating set theory allows ExEm to exhibit significant advantage over baselines for the task of node classification on BlogCatalog.  ExEm strengthens the performance by $115.23 \%$, $61.90 \%$, $60.71 \%$ and $53.83 \%$ compared with
SDNE, Line, DeepWalk and Node2vec on Micro-F1 metric by considering $80 \%$ data as training. Moreover, ExEm shows $53.51 \%$, $35.69 \%$, $30.81 \%$ and $28.71 \%$ performance gains than SDNE, Line, DeepWalk and Node2vec based on Macro-F1 quality with the same amount of training data. These results indicate how effective ExEm variation models are on BlogCatalog which is a denser network than PPI. Node2vec and DeepWalk follow a similar trend and both of them outperform Line and SDNE, similar observations to PPI. In contrast to PPI, Node2vec works better than DeepWalk with a gain of $0.34 \%$ and $3.73 \%$ with regards to Micro-F1 and Macro-F1. And Line operates more effectively than SDNE. 

--\textbf{{Wikipedia dataset}}: The result shows that there is an improvement between the results of ExEm and other methods on  Wikipedia dataset specifically in terms of Macro-F1.  ExEm acquires benefits of $18.99 \%$, $16.75 \%$, $15.34 \%$ and $13.32 \%$ comparing to  SDNE, Line, DeepWalk and Node2vec on Micro-F1 by selecting  $10 \%$ of nodes for training. Also, we have seen that ExEm boosts the efficiency by $59.75 \%$, $48.86 \%$, $48.48 \%$ and $36.93 \%$ percents above SDNE, Line, DeepWalk and Node2vec, respectively, for Macro-F1 score. These outcomes are as evidence to imply the potential of our random walk based method to represent Wikipedia's network structure better, which is also a dense word co-occurrence network \citep{qiu2018network}, comparing to the baselines. Selecting the best values for Node2vec parameters evinces this method outperforms DeepWalk. Still the performance of SDNE is the worst among the graph embedding techniques in this case.

--\textbf{{Scopus dataset}}: 
As can be seen from the results, ExEm obtains a great improvement in performance over the classification task on Scopus dataset. ExEm enhances the performance, given $80 \%$ amount of training data, about  $5.82 \%$, $2.80 \%$ and $2.52 \%$  over Line, DeepWalk and Node2vec in terms of Micro-F1 score.  For Macro-F1 metric,  the gains obtained by ExEm over these three baselines are $15.88 \%$, $8.30 \%$ and $10.07 \%$ , individually. We have made three observations on obtained results from Scopus dataset. Firstly, it is obvious that no results are presented for SDNE. The reason is the SDNE's prohibitive memory necessities for the input adjacency matrix. In other words, SDNE could only be run for smaller graphs and it fails to finish successfully for large graphs such as Scopus. Secondly, since Scopus network has the highest density in comparison to three other datasets, it has the largest values of Micro-F1 and Macro-F1 scores. Thirdly, we used a trial-and-error procedure in the selection of Node2vec parameters as its first running on Scopus.  Despite DeepWalk and Node2vec generate rather similar outcomes based on Micor-F1 score, DeepWalk is superior to Node2vec in terms of Macro-F1.

\subsubsection{Effect of dimension}
\begin{figure}
	\centering
	\includegraphics[width=0.9\textwidth]{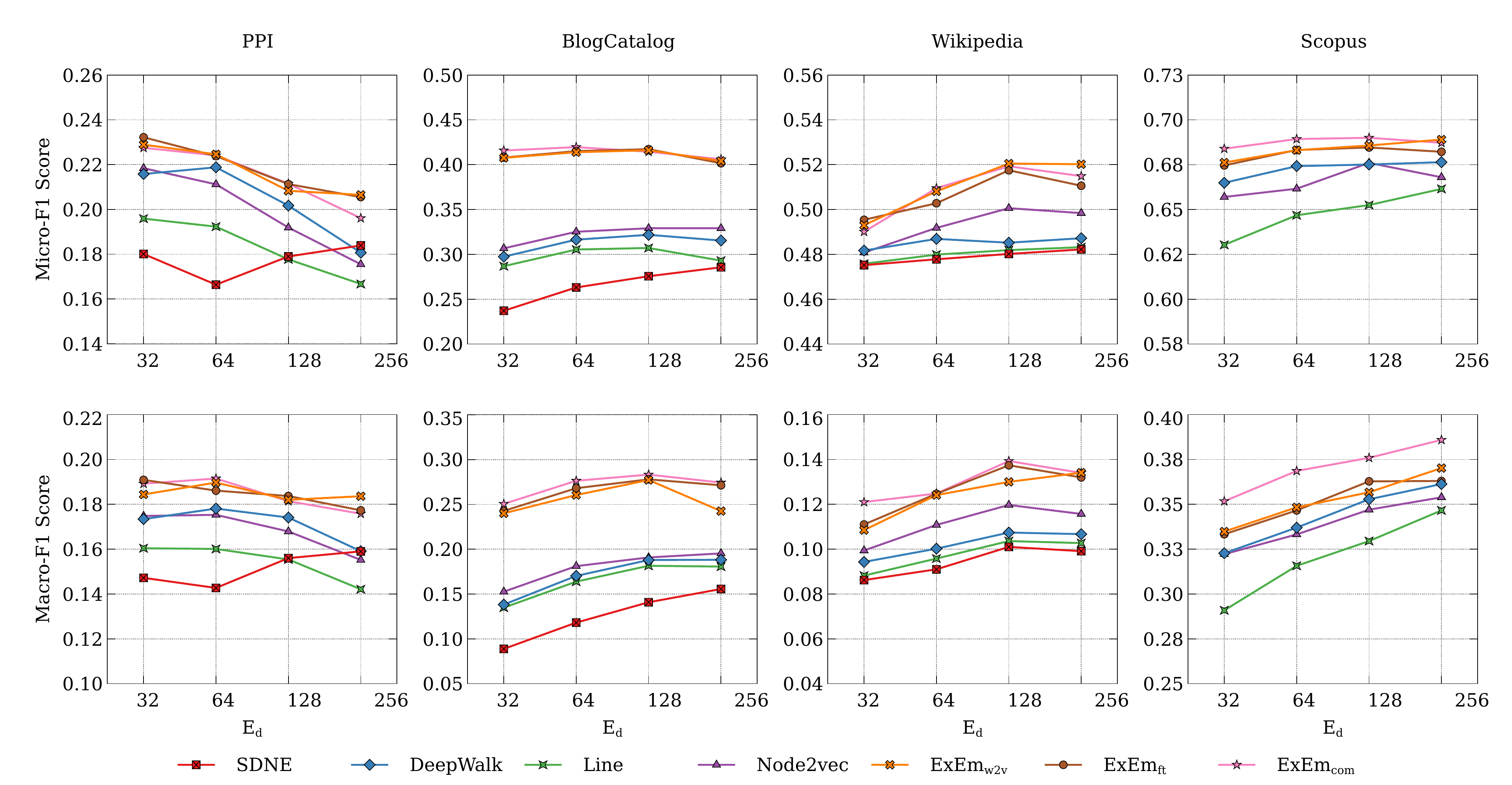}
	\caption{Micro-F1 and Macro-F1 scores on node classification task for different datasets over the diverse number of dimensions (dimensions of ExEm$_{com}$ are $64$, $128$, $256$ and $512$ and the train ratio is $50 \%$).} 
	\label{multi-label-calssification-di-re}
\end{figure}

Also, we studied the effect of embedding dimensions on node classification task for different approaches. We conducted the investigations by following the same experimental procedure done for different train ratios, with a change that we fixed the train ratio with a value of $50 \%$. It should be noted that the dimensions of ExEm$_{com}$ in this experiment equal $64$, $128$, $256$ and $512$ with regard to the embedding vector sizes  $32$, $64$, $128$ and $256$ for both ExEm$_{ft}$ and ExEm$_{w2v}$. Figure \ref{multi-label-calssification-di-re} illustrates the impacts of different embedding dimension sizes on various graph embedding approaches. The observations from the results lead to the conclusion that although the performance of all these graph embedding techniques go up gradually over the train ratios in the most datasets, we saw the uptrend and downtrend or sideways trends in the performance of techniques by varying the number of dimensions. The reason is that despite high-dimensional embedding presents more features of nodes, in some cases using a large dimensionality results in overfitting.  In PPI dataset, the performance of all methods with the exception of SDNE degrades as the number of dimensions increases. While SDNE's performance enhances as embedding dimension increases above $128$, ExEm achieves the best performance on PPI with $32$ and $64$ dimensions for ExEm$_{ft}$ and ExEm$_{w2v}$, and ExEm$_{com}$, respectively.  With a couple of exceptions, Micro-F1 and Macro-F1 scores increase as embedding dimension increases in BlogCatalog, Wikipedia and Scopus datasets. Also, it appears that ExEm outperforms other methods.  ExEm$_{ft}$ and ExEm$_{w2v}$, and ExEm$_{com}$ are able to embed nodes to vectors with $128$ and $256$ dimensions, correspondingly, with high scores over all datasets. Among different forms of ExEm, ExEm$_{w2v}$' results are closer to ExEm$_{ft}$ except in a few cases. Due to the drawback of SDNE to operate over large networks, no result is reported for it on Scopus. 

\subsection{Link prediction}\label{Link_prediction_re}
By taking the learned node representations as inputs, we accomplish the link prediction experiment to compare the effectiveness of ExEm method with four other approaches.  As mentioned before, link prediction can be treated as a binary classification task where the objective function is defined by the AUC score. After obtaining embeddings that are $128$-dimensional vectors, we randomly hid $50 \%$ of the network edges for each dataset. Then, we predicted the existence or non-existence of links between pairs of nodes in the rest of network by training a Logistic Regression classifier. To guarantee a reasonable examination, we repeated the prediction process for $10$ times and reported the mean value of AUC score. Moreover, to provide an edge representation for node pair $(u,v)$, we extended algorithms with different binary operators. These operators are defined by the following equations \citep{KEIKHA201847,grover2016Node2vec,crichton2018neural}:

\begin{equation}\label{heuristic_scores_av}
	Average  = \frac{V(u)_{i} + V(v)_{i}}{2}
\end{equation}

\begin{equation}\label{heuristic_scores_ha}
	Hadamard  = V(u)_{i} \times V(v)_{i}
\end{equation}

\begin{equation}\label{heuristic_scores_l1}
	Weighted-L1  = \mid V(u)_{i} - V(v)_{i} \mid
\end{equation}

\begin{equation}\label{heuristic_scores_l2}
	Weighted-L2  = \mid V(u)_{i} - V(v)_{i} \mid ^ 2
\end{equation}

where $V(u)_{i}$ and $V(v)_{j}$ are the $i$th features of $u$ and $v$, respectively.

\subsubsection{Results}
Table \ref{Link_prediction_re_tb} shows the summarized results of the AUC score for different methods on the task of link prediction over four datasets. According to these results we have the following observations: \begin{enumerate*}[label={\roman*)},font={\bfseries}]
	\item LINE and SDNE are blamed for their poor performance in link prediction, as they can not find the pattern of edge existence in graphs.
	\item  DeepWalk and Node2vec perform better than LINE and SDNE because of employing the random walk based model which can better obtain proximity information within nodes.
	\item By large margins the improvement of ExEm over the  baselines is more obvious in link  prediction task. ExEm promotes the efficiency of link prediction on Weighted$-L2$ operator about $55.53 \%$, $45.49 \%$ , $28.63 \%$ , and $25.59 \%$ over SDNE, Line, DeepWalk and Node2vec, accordingly.
	Based on the obtained results of average operator from Astro-PH graph, we see the $22.26 \%$, $26.35 \%$ , $19.99 \%$ , and $16.93 \%$  improvements of ExEm than SDNE, Line, DeepWalk and Node2vec, respectively.  ExEm improves AUC scores on BlogCatalog by $19.80 \%$, $53.20 \%$ , $27.88 \%$ , and $41.40 \%$  over SDNE, Line, DeepWalk and Node2vec for Weighted$-L1$ operator. For Scopus dataset, ExEm achieves gains of $23.91 \%$ , $29.85 \%$ , and $21.04 \%$ in comparison with Line, DeepWalk and Node2vec for average operator and also there is no result for SDNE due to its inability to operate on large network. Our explanation for the performance of ExEm on link prediction task is that each node in the network has at least one neighbor of dominating nodes which effectively dominate the connections of nodes in a network, so ExEm can predict the most likely edges which are not observed in the training data from the learned embedding. The comparison of different models of ExEm presents that ExEm$_{com}$ reveals a better performance than two other forms After ExEm$_{com}$, ExEm$_{w2v}$ gains the second place on all datasets.
\end{enumerate*}

\begin{table}
	\caption{AUC score of link prediction for different datasets on various operators (dimension of ExEm$_{com}$ is $256$). (a)Average, (b)Hadamard, (c)Weighted$-L1$, and (d)Weighted$-L2$. }
	\label{Link_prediction_re_tb}
	\centering
	\begin{tabu}{c| c| c|c|c|c}
		\hline
		Op &  Algorithm & \multicolumn{4}{c}{Dataset}\\
		\cline{3-6}
		& & PPI & Astro-PH & BlogCatalog & Scopus \\
		\hline
		&		SDNE&						    0.6782&		        0.6262&         0.6965&         -\\			
		&		LINE&						    0.6424&				0.6059&         0.7829&         0.5398\\
		&		DeepWalk&					    0.627& 			    0.638&          0.7636&         0.5151\\
		&		Node2vec&				    	0.7543& 			0.6547&         0.7493&         0.5526\\
		(a) &	ExEm$_{ft}$& 		                0.8041&		        0.7636&         0.7829&         0.6661\\
		&		ExEm$_{w2v}$&		                0.8034&				0.7612&         0.7976&         0.664\\
		&	    ExEm$_{com}$&	            \textbf{0.8098}& 	\textbf{0.7656}& \textbf{0.7999}& \textbf{0.6689}\\
		\hline
		&		SDNE&					        0.6981&		        0.7117&         0.66&           -\\	
		&		LINE&					    	0.7314&		        0.9352&         0.7766&         0.8364\\
		&		DeepWalk&				    	0.7441&		        0.9335&         0.7256&         0.9607\\
		&		Node2vec&				    	0.7719&		        0.9583&         0.7632&         0.9693\\
		(b) &	ExEm$_{ft}$& 		                0.9278&	            0.9765&         0.8041&         0.9874\\
		&		ExEm$_{w2v}$&		                0.9262&		        0.9766&         0.8026&         0.9875\\
		&		ExEm$_{com}$&             \textbf{0.9454}& 	\textbf{0.983}&  \textbf{0.8335}& \textbf{0.9908}\\
		\hline
		&		SDNE&						    0.6436&		        0.6066&         0.6001&         -\\
		&		LINE&					    	0.6796&		        0.8948&         0.7674&         0.8428\\	
		&		DeepWalk&				    	0.8753&		        0.8966&         0.7189&         0.9656\\
		&		Node2vec&				    	0.6292&		        0.9132&         0.6502&         0.975\\
		(c) &	ExEm$_{ft}$& 		                0.9657&	    \textbf{0.9886}&        0.9078&         0.9929 \\
		&		ExEm$_{w2v}$&		                0.971&	    	    0.988&          0.9141&         0.9934\\
		&		ExEm$_{com}$&             \textbf{0.9726}&    	    0.9876& \textbf{0.9194}&  \textbf{0.9935}\\
		\hline
		&		SDNE&						    0.636&		        0.5761&         0.5978&         -\\
		&		LINE&						    0.6799&		        0.8932&         0.7507&         0.8304\\
		&		DeepWalk&					    0.6118& 	        0.8981&         0.7234&         0.9864\\
		&		Node2vec&					    0.6236& 	        0.9146&         0.6529&         0.9757\\
		(d) &	ExEm$_{ft}$& 		                0.9708&		\textbf{0.9892}&        0.9137&         0.9928\\
		&		ExEm$_{w2v}$&		                0.9749&		        0.985&          0.9207&         0.9929\\
		&		ExEm$_{com}$&             \textbf{0.9753}& 	\textbf{0.9892}& \textbf{0.9212}&  \textbf{0.9938}\\
		\hline
	\end{tabu}
\end{table}

\subsubsection{Effect of dimension}
\begin{figure}
	\begin{center}
		\centering
		\includegraphics[width=1\textwidth]{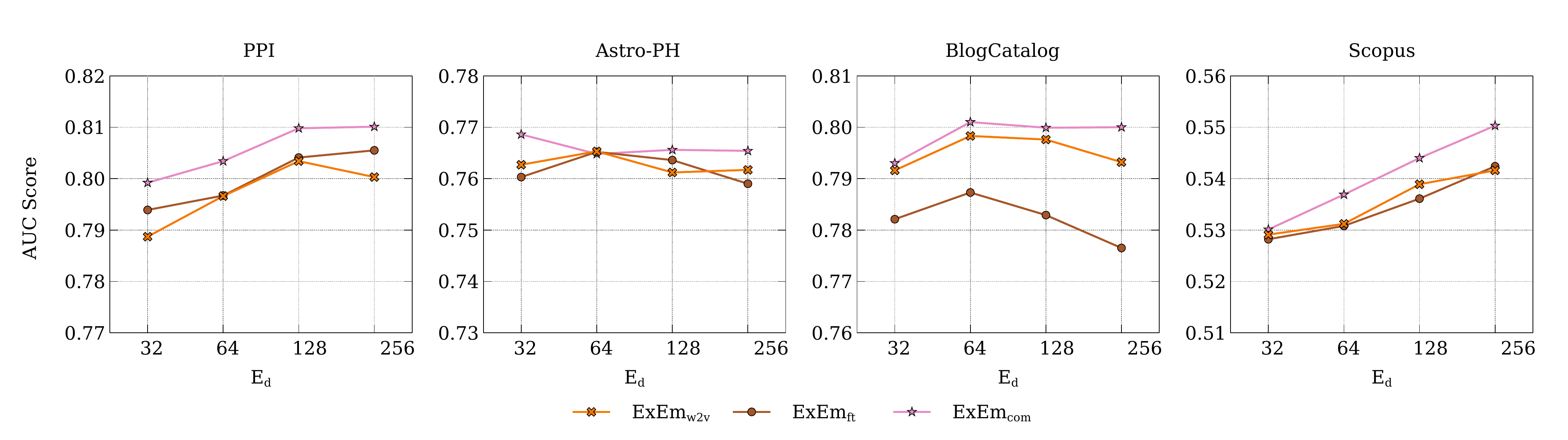}
		\caption{AUC score of link prediction for different datasets with varying dimensions on the average operator (dimensions of ExEm$_{com}$ are $64$, $128$, $256$ and $512$).}
		\label{link-pre-di-re} 
	\end{center}
\end{figure}

Additionally, we investigated the effect of embedding dimensions on only different methods of ExEm in the link prediction task. We followed the same strategy as mentioned above, just using differed dimension sizes  and the average operator to provide more insights on the performance of ExEm. Figure \ref{link-pre-di-re} illustrates the effect of embedding dimensions on ExEm models.  Overall, the AUC score increases over the dimension given. As with node classification, we observed that ExEm$_{ft}$ and ExEm$_{w2v}$, and ExEm$_{com}$ achieve the best performance on all datasets with $128$ and $256$ dimensions, respectively. Based on the results, ExEm$_{com}$ outperforms ExEm$_{ft}$ and ExEm$_{w2v}$ since the higher number of dimension makes it capable of storing more information. Also, we found that ExEm$_{ft}$ and ExEm$_{w2v}$ show the same trends by increasing the size of node embeddings. In BlogCatalog and Scopus, ExEm$_{w2v}$ is the winner, while ExEm$_{ft}$ overcomes ExEm$_{w2v}$ in PPI and Astro-PH.

\subsection{Recommendation}\label{case-study-result}
 The purpose of this experiment is to show how a graph embedding approach can be effectively used to order item recommendations with the help of the learned node embeddings. As previously described, this paper introduces a novel strategy for computing experts' scores using the expert embedding vectors and recommending top experts whose scores are high. So, we conducted a case study to demonstrate the efficacy of ExEm in the recommendation task. We selected three topics: information extraction (IE), natural language processing (NLP), and machine learning (ML) from \textit{Arnetminer} data. The lists of people in these topics are used as experts to construct the ground truth to evaluate  the recommendation task on the Scopus dataset. Note that our task is not to predict the exact score value of each expert but to rank them in terms of their positions in the list. That means we take into account the position of the experts in these lists as their ranks for the ground truth. We used cosine similarity to measure the distance between the node embedding vectors and centroid. We recommended the nearest nodes to the centroid as experts. The dimension of expert vectors is fixed to $128$. We announced the results in terms of nDCG$@k$. In using nDCG$@k$, we set $k$ to $5$, $10$ and $15$. Because of the weakness of SDNE to run on large dataset, we compared ExEm with Line, DeepWalk, Node2vec approaches.

\subsubsection{Results}
Table \ref{recommendation_re_tb} demonstrates nDCG score provided by the identified top $k$ experts in three specific topics. As can be seen, except in a few cases, ExEm$_{ft}$ has gained the highest values among the competitors. Then,  ExEm$_{com}$ takes the second-ranking position and provides better performance in comparison to ExEm$_{w2v}$, Node2vec, DeepWalk and Line. Also, it is clear that unlike node classification and link prediction tasks, Line shows comparable performance to Node2vec, and DeepWalk performs poorly. Over and above that we compared ExEm against of ExpertiseRank \citep{zhang2007expertise} in order to compare its performance with studies in the expert recommendation system domain. ExpertiseRank is  a  PageRank-like  algorithm used to calculate experts' score in the user-user graph based on ask-answer relations of the users. ExpertiseRank considers the number and quality of connections  of a candidate expert to determine a rough estimate of how important the candidate is. It is clear that ExEm outperformed the ExpertiseRank  in all three topics. The reason is that ExpertiseRank  tries to find experts based on the degree of connections of experts with others in the collaborative network. While ExEm raises its awareness of experts' expertise through their embeddings which present rich information about experts. The other explanation for the success of our proposed method is its way of calculating expert scores.  

In summary, we provided two important feedbacks  form the results. Primarily, the high values of nDCG scores for graph embedding methods in comparison with ExpertiseRank  show that our introduced strategy provides an efficient solution for computing experts' scores based on expert embeddings.  In addition, as ExEm generates more appropriate embeddings for experts of different topics than comparative baselines, using the expert embeddings obtained by ExEm models specially ExEm$_{ft}$ makes significant gains in the expert recommendation system.

\begin{table}[]
	\caption{nDCG score of recommendation for Scopus dataset based on top k experts (dimension of ExEm$_{com}$ is $256$). }
	\label{recommendation_re_tb}
	\begin{tabular}{l|c|c|c|c|c|c|c|c|c}
		\hline
		Topics                         & \multicolumn{3}{c|}{ML}                             & \multicolumn{3}{c|}{NLP}                            & \multicolumn{3}{c}{IE}                             \\ \hline
		nDCG$@$                          & $5$           & $10$        & $15$        & $5$           & $10$        & $15$        & $5$           & $10$        & $15$        \\ \hline
		Line&           0.5005&             0.6137&             0.7218&             0.6112&             0.6295&             0.6032&             0.4052&         0.5028&             0.6089\\ \hline
		DeepWalk&       0.3119&             0.4183&             0.6156&             0.2720&             0.4268&             0.4329&             0.4666&         0.4898&             0.6203\\ \hline
		Node2vec&       0.5286&             0.5897&             0.7055&             0.5124&             0.5570&             0.5971&             0.4865&         0.5382&             0.6361\\ \hline
		ExpertiseRank&       0.4924&             0.5796&             0.6646&             0.5477&             0.5539&             0.6036&             0.5200&         0.5467&             0.6799 \\ \hline     
		ExEm$_{w2v}$&       0.6455&             0.6415&             0.8374&             \textbf{0.6243}&    0.6486&             \textbf{0.6538}&    0.5734&         0.5602&             0.7089\\ \hline
		ExEm$_{ft}$&        \textbf{0.7029}&    \textbf{0.6462}&    \textbf{0.8428}&     0.6239&             0.6482&             0.6403&             \textbf{0.5882}& \textbf{0.5724}&   \textbf{0.7202}\\ \hline
		ExEm$_{com}$&      0.7020&             0.6459&             0.8414&             0.6240&             \textbf{0.6494}&    0.6411&             0.5747&          0.5719&            0.7114\\ \hline
	\end{tabular}
\end{table}

\subsubsection{Effect of dimension}
\begin{figure}
	\begin{center}
		\centering
		\includegraphics[width=0.3\textwidth]{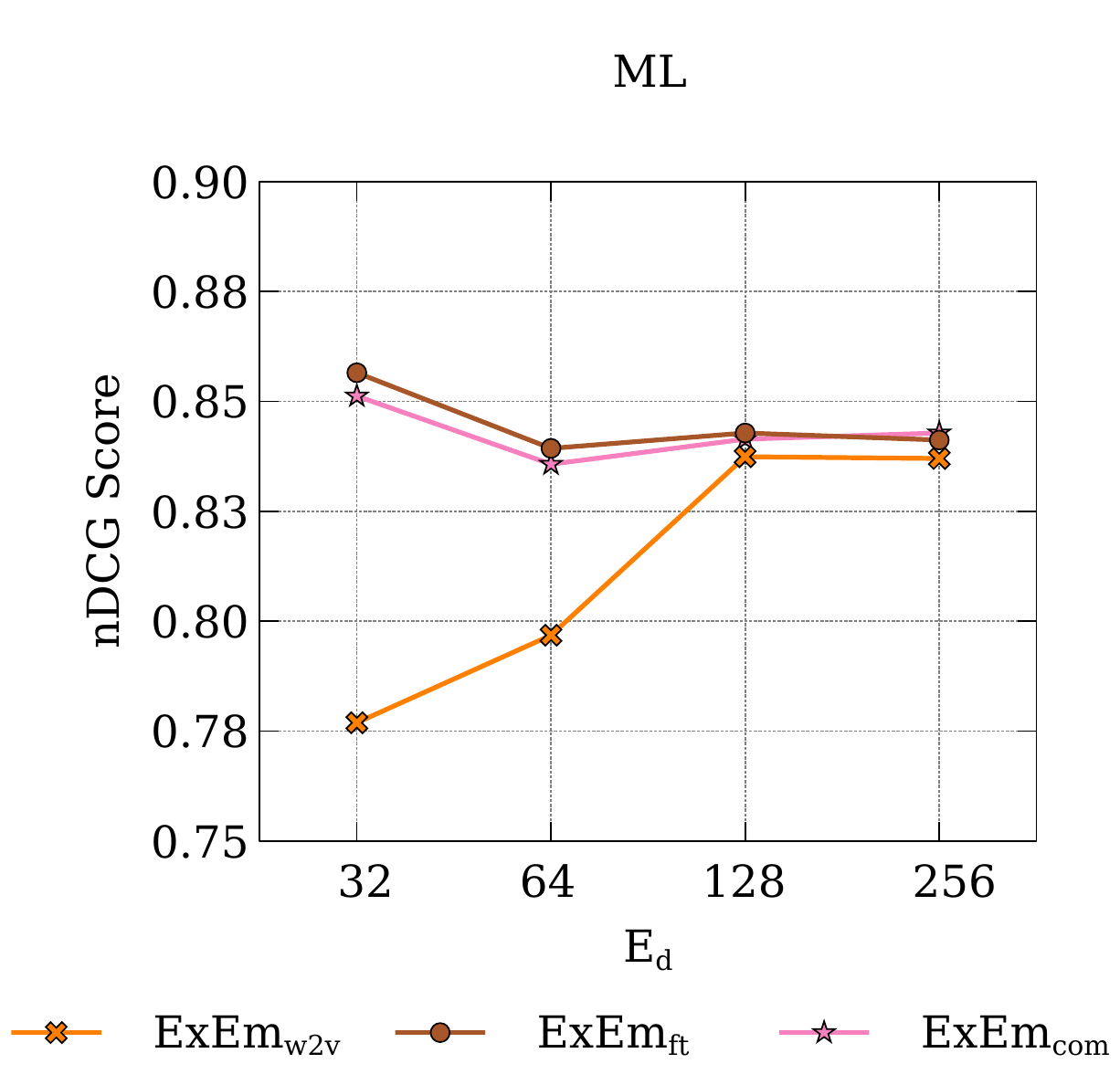}
		\caption{nDCG$@15$ score of recommendation for ML topic with varying dimensions (dimensions of ExEm$_{com}$ are $64$, $128$, $256$ and $512$).}
		\label{recomm-di-re} 
	\end{center}
\end{figure}

Also, we explored the effect of embedding dimensions on only different methods of ExEm for recommendation task. We used the same strategy as mentioned before, by merely limiting our test into ML topic and nDCG$@15$. Figure \ref{recomm-di-re} illustrates the effect of embedding dimensions on ExEm models in our case study. It is clear that although ExEm$_{ft}$ outperforms, it reveals an identical trend to ExEm$_{com}$. The performances of both ExEm$_{ft}$ and ExEm$_{com}$ decline with a small slope at the beginning and then their performances saturate as the number of dimensions increase. However, we see that the performance of ExEm$_{w2v}$ initially increases slightly faster, but it finally shows a fixed-performance like two other methods with the increase in the size of expert embeddings.

\subsection{Parameter sensitivity}\label{parameter_sen_sec}
As mentioned before, there exist three common ways to combine the features obtained from fastText and Word2ve and create a single representation for each node. So we examine how different choices of merging features affect the performance of ExEm. For this evaluation, we measured the Micro-F1 and Macro-F1 scores for the node classification task on the BlogCatalog dataset using  $10 \%$ to $90 \%$ splits between labeled and unlabeled nodes with embedding size $128$.  Besides, ExEm involves a number of parameter that may effect its performance. Therefore, we conduct a sensitivity analysis of ExEm to context window size $w$ and length of random walks $L_R$ parameters. For sensitivity investigation we followed the first test setting just using $50 \%$ as training data and the remaining as test data.

\subsubsection{Results}
\begin{figure}
	\begin{center}
		\centering
		\includegraphics[width=0.9\textwidth]{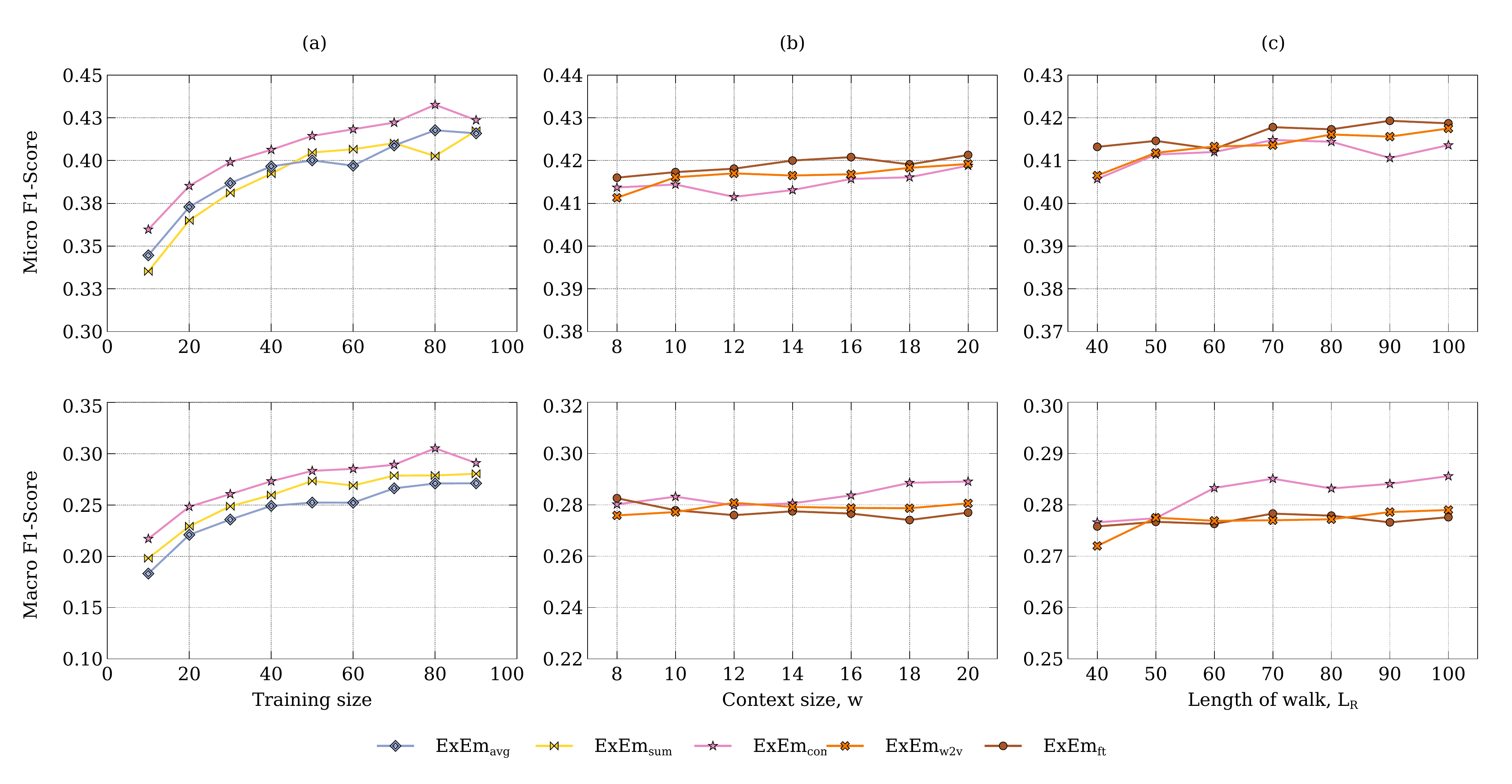}
		\caption{(a) Analysis of different combinations of the features obtained from fastText and Word2ve, (b) and (c) parameter sensitivity for node classification on the BlogCatalog network.}
		\label{parameter_sen}
	\end{center}
\end{figure}

As we can see in Figure \ref{parameter_sen}a, ExEm$_{com}$ consistently and significantly outperforms ExEm$_{sum}$ and ExEm$_{avg}$ in terms of both metrics. This increase in performance can be based on concatenation function that conducts the dimension of node vector space becomes higher, and so ExEm$_{com}$ can preserve most of the meaningful information about nodes without altering data. Although summing and averaging reduce node embedding size, they lose some information and hence they perform poorly. Moreover, according to Micro score, ExEm$_{avg}$ works better than ExEm$_{sum}$, while Macro score shows different deduction.

Moreover, Figures \ref{parameter_sen}b and \ref{parameter_sen}c suggest that context window size and length of random walks are positive to the node classification performance. However, 
they have relatively little relevance to the performance and the differences are not that large in these cases. Briefly, according to the analysis, various models of ExEm are not strictly sensitive to these parameters and can achieve high performance under a affordable parameter choice. 

\section{Discussion}\label{discussion}
It can be inferred from the results that Line as an edge modeling based approach which uses first- and second-order proximities and SDNE as a deep learning based method that employs auto-encoders do not make any additional gains as compared to random walk based methods in all tasks. It can be seen that in all instances except for PPI network over node classification, SDNE performs poorly. This is because SDNE focuses on the homophily objective to map the connected node pairs closer to each other and ignores the structural roles of nodes. Also, the other drawback of SDNE is that it is not scalable to large  graphs such as Scopus because of its memory consumption in order to feed the complete adjacency nodes as inputs. By the same token, Line embeds nodes closer which share common $one-$hop neighborhood, while it does not pay attention to their roles.  On the other hand,  random walk based graph embedding methods, ExEm, DeepWalk and Node2vec, show promising results over node classification, link prediction and recommendation tasks. Since random walks can tend to spread quickly over a local area, they can better capture local community structure and concurrently investigate different parts of the same graph \citep{perozzi2014deepwalk}. Additionally, we  observed that Node2vec and DeepWalk outperform Line and SDNE especially in link prediction and node classification tasks, although there are a number of problems with them that are solved by ExEm through using dominating set. One of the issues that DeepWalk encounters is its randomness which provokes Deepwalk not to preserve the local neighborhood of the node well and  makes a lot of noises mostly for nodes with high degrees \citep{wang2016structural}. Another drawback  of DeepWalks is that it does not embed nodes from this outlook that nodes with similar roles should be embedded closely together. However, Node2vec proposes a biased random walk that addresses the problems related to DeepWalk by virtue of two arguments $p$ and $q$. The common problem of Node2vec is that these arguments should be valid for a certain set of values for each network in order to properly produce node representations that take into consideration the homophily and structural equivalence assumptions. Therefore, because of the dependency of Node2vec's performance on adjusting the values of these parameters, we see that in some cases DeepWalk performs well compared to Node2vec. For example, we set parameters as $p=0.25$, $q=2$ and $p=0.5$, $q=4$ for PPI and BlogCatalog networks, respectively. For PPI dataset those values are the worst choices and the outcome has the poor performance of Node2vec than DeepWalk in node classification task, while in Blogcatalog the parameters put Node2vec in the second highest Macro-F1 and Micro-F1 scores after various versions of ExEm. Thus, the values of these parameters must be carefully chosen for each network to achieve a good performance. Based on the observations we found out that ExEm is more robust and effective technique for capturing node representations  on  all test  graphs. Taking advantage of dominating nodes in random walks helps ExEm to work efficiently on a variety of networks including large and dense graphs like Scopus or BlogCatalog. The reason is that the virtual backbone formed by dominating nodes can efficiently control the structure of graph and retrieve information from it \citep{Xthesis}. Moreover, having the second dominating node in the walk makes the connections between different parts of a graph. In other words, the attendance of the first and second dominating nodes encourages ExEm to obey homophily and structural role equivalences in encoding nodes and provides ExEm a higher learning flexibility than baselines. In brief, the main  differences between ExEm and the other methods are: \begin{enumerate*}[label={\roman*)},font={\bfseries}]
	\item ExEm uses an intelligent random walk sampling strategy which is based on dominating nodes.
	\item ExEm  is more effective than Line, SDNE, DeepWalk and Node2vec, as is illustrated by  our  experiments  in three  different tasks on various graphs. \item ExEm  is  efficient  for  dense graphs  and scalable  for  very  large  applications. 
	\item ExEm has the lowest execution time among both DeepWalk and Node2vec since ExEm's intelligent random walk starts from only dominating nodes instead of all nodes. Also, the second reason is that the second dominating node exists in the rest of the walk with probability values $0.44$, $0.33$, $0.50$ and $0.30$ and $0.27$ obtained from experiments over different datasets PPI, Wikipedia, BlogCatalog, arXiv(Astro-PH) and Scopus, respectively. Hence, it is not necessary to investigate the expression of second dominating node in each random walk. While the computation of transition probabilities for going from one node to another in Node2vec is taking more time to generate random walks. For instance, we calculated the execution time of ExEm$_{w2v}$, DeepWalk and Node2vec on Blogcatalog for node classification task and the results show that DeepWalk learns node representations in $114.62$ seconds which is faster than Node2vec with runtime $294.94$ seconds. We found that in ExEm the time of finding a dominating set and generating random walks equal to $0.039$ and $28.19$ seconds, respectively. By adding the training time, the total execution time of ExEm$_{w2v}$ is $106.439$ seconds which is shorter than Node2vec and DeepWalk.  
	\item ExEm can easily accommodate itself to dynamic networks only by adding new random walks from the changed part, while  Node2vec, Line and SDNE can not cope with dynamic graphs.
\end{enumerate*}

Besides, we also note that our proposed scheme for estimating experts' scores based on expert embeddings addresses the issue of expert finding in a social network.  Using expert embeddings created by ExEm in the proposed method significantly outperforms all works to rank candidate experts and recommend top experts accurately. In addition, we highlighted the fact that almost all methods conducted on our collected dataset, Scopus, are better than the experiments conduced on other datasets. One of the reasons is higher density of Scopus compared to other datasets. 

\section{Answers to Research Questions}\label{ans-que}
In what follows, we are going to answer the research questions from Section \ref{sec:introduction} based on the observations from extensive experimental comparison:

\textbf{\ref{rq1}}
The results proved the advantage of our collected dataset for different usages. The value of the modularity shows the efficiency of Scopus data for community detection task. Moreover, the values of various scores obtained from conducting graph embedding techniques on Scopus graph underline the usefulness of this dataset for multi-label classification, link prediction and recommendation tasks.

\textbf{\ref{rq2}}
Experimental results demonstrated that creating intelligent random walks by using dominating nodes not only declines runtime, but also provides key insight into the organization of network. ExEm hires two dominating nodes in each  path  sampled to simultaneously preserve the local and global network structures. The first dominating node characterizes the local neighborhoods accurately, while the second dominating node helps ExEm to learn the node embeddings based on their similar structural roles within the network topology.

\textbf{\ref{rq3}}
We proposed a novel strategy that computes experts' scores based on the expert embedding vectors and accurately recommends experts. The proposed method extracts experts whose subject areas include the given topic and makes a cluster by them. Then, the center of this cluster is found by taking the average of all the expert embedding vectors in the group. Then, cosine similarity  measures the distance between the embedding vectors and centroid. Finally, the nearest nodes to the centroid are recommended as experts.  We observed that using expert embeddings created by ExEm in the proposed method significantly outperforms all works to rank candidate experts. Note that this approach can be applied to any types of graph with a special example of the graph related to the relationship between questioner and answerer in $QAC$s such as \textit{StackOverflow} and \textit{Quara}.

\section{Conclusion} \label{sec:conclusion}
In this paper, we have proposed two approaches and presented a new dataset. Our first proposed approach is a random walk based graph embedding technique, called ExEm, that incorporates the dominating set from graph theory to graph embedding. Starting random walks with dominating nodes and existing another dominating node in the following of each sampled path help ExEm to fulfill homophily and structural role objectives. ExEm uses three embedding methods including Word2vec, fastText and the concatenation of these two to extract node embeddings from these random walks. Experimental results demonstrated that ExEm is significantly more effective and applicable than SDNE, Line, DeepWalk and Node2vec over multi-label classification, link prediction and recommendation tasks. Also, this research represented another approach used to compute experts’ scores based on expert embedding vectors. This proposed framework achieved much better performance than ExpertiseRank approach in the recommendations of top experts. Finally, we presented a dataset related to a co-author network formed by crawling the vast author profiles from Scopus.




\bibliographystyle{apa}
\bibliography{sample}

\end{document}